\documentclass[12pt]{iopart}
\usepackage{iopams}
\newfont{\myeu}{eurm10 at 12pt}
\newfont{\bfrak}{eufb10 at 12pt}
\def\wb{{\overline W}}
\def\bZ{{\bf Z}}

\def\bfB{{\mathbf{\bar B}}}
\def\bfC{{\mathbf{\bar C}}}
\def\bfE{{\mathbf{\bar E}}}
\def\bfH{{\mathbf{\bar H}}}
\def\bE{{\bar E}}

\def\bI{{\bar I}}
\def\bN{{\bar N}}
\def\bK{{\bar K}}
\def\vp{^{\vphantom{i}}}
\def\vn{_{\vphantom{i}}}
\def\vf{{\vphantom{k^k}}}
\def\half{\textstyle{\frac12}}
\def\halfs{\scriptstyle{\frac12}}
\def\sfactor#1#2{\Bigg[\begin{array}{@{}c@{}}#1\\#2\end{array}\Bigg]}
\begin{document}
\title[Quantum Loop Subalgebra]%
{Quantum Loop Subalgebra and Eigenvectors of the Superintegrable Chiral
Potts Transfer Matrices}

\author{Helen Au-Yang$^{1,2}$ and Jacques H H Perk$^{1,2}$%
\footnote{Supported in part by the National Science Foundation
under grant PHY-07-58139 and by the Australian Research Council
under Project ID: LX0989627.}}

\address{$^1$ Department of Physics, Oklahoma State University, %\\
145 Physical Sciences, Stillwater, OK 74078-3072, USA%
\footnote{Permanent address.}\\
$^2$ Centre for Mathematics and its Applications \&\
Department of Theoretical Physics,
Australian National University, Canberra, ACT 2600, Australia}
\ead{perk@okstate.edu, helenperk@yahoo.com}
\begin{abstract}
It has been shown in earlier works that for $Q=0$ and $L$ a multiple
of $N$, the ground state sector eigenspace of the superintegrable
$\tau_2(t_q)$ model is highly degenerate and is generated by a
quantum loop algebra $L({\mathfrak{sl}}_2)$. Furthermore, this loop
algebra can be decomposed into $r=(N\!-\!1)L/N$ simple
${\mathfrak{sl}}_2$ algebras. For $Q\ne0$, we shall show here that the
corresponding eigenspace of $\tau_2(t_q)$ is still highly degenerate,
but splits into two spaces, each containing $2^{r-1}$ independent
eigenvectors. The generators for the ${\mathfrak{sl}}_2$
subalgebras, and also for the quantum loop subalgebra, are given
generalizing those in the $Q=0$ case. However, the Serre relations
for the generators of the loop subalgebra are only proven for some
states, tested on small systems and conjectured otherwise.
Assuming their validity we construct the eigenvectors of
the $Q\ne0$ ground state sectors
for the transfer matrix of the superintegrable chiral Potts model.
\end{abstract}

%%%%%%%%%%%%%%%%%%%%%%%%%%%%%%%%%%%%%%%%%%%%%%%%%%%%%%%%%%%%%%%%%%%%%%%%
\section{Introduction}

Since the introduction of the integrable chiral Potts model
\cite{AMPTY,BPAuY88}, with Boltzmann weights parametrized by a curve of
genus $g>1$ and satisfying the star-triangle equation, there has been
a lot of progress. Much insight can be gained by studying the
superintegrable subcase which has a representation of the Onsager
algebra built in and whose associated uniform $N$-state quantum
chain was discovered in 1985 \cite{vonGehlen1985}.

Solving for the free energy of the $N$-state superintegrable chiral
Potts model on an $L\times\infty$ face-centered square lattice
with periodic boundary conditions and in the commensurate phase,
Baxter \cite{Baxsu,BaxIf1,BaxIf2} discovered a special set of
$2_{\vp}^{m_Q}$ eigenvalues of the transfer matrix expressed in
terms of the roots of the Drinfeld polynomial
\begin{equation}
\fl P_Q(z)=N^{-1}t^{-Q}\sum_{a=0}^{N-1}
\omega^{-Qa}\frac{(1-t^N)^L}{(1-\omega^a t)^L}=\sum_{m=0}^{m_Q}
\Lambda^{Q}_{ m}z^m
=\Lambda^{Q}_{ m_Q}\prod_{j=1}^{m_Q}(z-z_{j,Q}),
\label{roots}
\end{equation}
with
\begin{equation}
z\equiv t^N,\quad m_Q\equiv\lfloor L(N-1)/N-Q/N\rfloor.
\end{equation}
For $0\le Q\le N-1$, $\omega^Q$ denotes the eigenvalue of the
spin shift operator ${\cal X}$, shifting all spins in a row by one.
The eigenspace associated with the special $2_{\vp}^{m_Q}$ eigenvalues
is called the $Q$ ``ground state sector," as one of them gives the
ground state energy of the superintegrable quantum chain in the
$Q$ sector.

To study the model in more detail, we will need explicit information
about eigenvectors. Such a study was initiated by Tarasov \cite{Tara},
who set up an algebraic Bethe Ansatz construction based on the
$\tau_2$ model, but did not address possible degeneracy in the
superintegrable $\tau_2$ transfer matrix eigenvalue spectrum. Even
though the $\tau_2$ and the chiral Potts transfer matrices commute,
eigenvectors of the $\tau_2$ model will typically fail to be
eigenvectors of the chiral Potts model, due to the degeneracy
in the $\tau_2$ spectrum.

For $Q\!=\!0$ and $L$ a multiple\footnote{Cases with $L$ not a multiple
of $N$ must be treated separately with methods as given for $Q\ne0$
in this paper. For the study of the thermodynamic limit $L\to\infty$,
however, we only need $L/N$ integer.} of $N$, it has indeed been shown
\cite{NiDe1,APsu1} that the ground state sector eigenspace
of $\tau_2(t_q)$ is highly degenerate, and that it supports a quantum
loop algebra $L({\mathfrak{sl}}_2)$. Furthermore, this loop algebra
can be decomposed into $r=m_0$ simple ${\mathfrak{sl}}_2$ algebras.
These results enabled us to express the chiral Potts transfer matrix
in terms of the generators of $r$ ${\mathfrak{sl}}_2$ algebras
\cite{APsu2}, so that the corresponding $2^r$ eigenvectors of the
transfer matrix were found, where $r=m_0=L(N-1)/N$. 

For $Q\ne0$ cases, some investigation for the six-vertex model at a root
of unity was done in \cite{DFM}; apart from that work not much more
was known explicitly. However, as the eigenvalues of transfer matrix
have exactly the same property for $Q=0$ as well as for $Q\ne0$, this
gave us confidence that it must work out somehow also for $Q\ne0$.
Here we report the progress that has been made. We generalized many of
the results that we obtained in \cite{APsu1,APsu2} for $Q=0$ to $Q\ne0$
cases by first checking these results on a computer for small $N$
and $L$ and then proving them analytically.

To obtain the eigenvectors of the superintegrable chiral Potts transfer
matrix outside the ground state sector one may start with the regular
Bethe vectors of the $\tau_2$ model \cite{Tara}, complete the
corresponding eigenvector sectors applying suitable quantum loop
algebras and choose suitable linear combinations in each sector
(as done in \cite{APsu2} starting from the ``ferromagnetic" state).
Partial progress along these lines has been reported \cite{NiDe2,Roan},
but no explicit results for chiral Potts eigenvectors were given.

Before proceeding, we will first discuss the differences between
our notations and those of Baxter for the $\tau_2(t_q)$
model \cite{Baxter-tau}, and with the work of Nishino and
Deguchi \cite{NiDe1}.

%%%%%%%%%%%%%%%%%%%%%%%%%%%%%%%%%%%%%%%%%%%%%%%%%%%%%%%%%%%%%%%%%%%%%%%%
\subsection{Preliminaries}

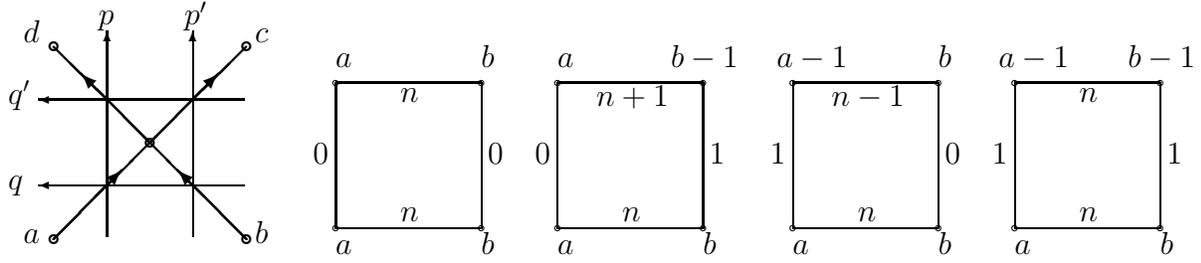
\begin{figure}[!hbtp]\begin{center}
\setlength{\unitlength}{0.57mm}
\begin{picture}(40,60)
\put(10,-2){\vector(0,1){48}}
\put(30,-2){\vector(0,1){48}}
\put(42,10){\vector(-1,0){48}}
\put(42,30){\vector(-1,0){48}}
\thicklines
\put(-2.5,-2.5){\line(1,1){24}}
\put(42.5,-2.5){\line(-1,1){24}}
\put(-2.5,-2.5){\vector(1,1){16}}
\put(42.5,-2.5){\vector(-1,1){16}}
\put(20,20){\line(-1,1){22.45}}
\put(20,20){\line(1,1){22.45}}
\put(20,20){\vector(-1,1){16}}
\put(20,20){\vector(1,1){16}}
\put(-2.5,-2.5){\circle{2}}
\put(20,20){\circle{2}}
\put(42.5,-2.5){\circle{2}}
\put(42.5,42.5){\circle{2}}
\put(-2.5,42.5){\circle{2}}
\put(-9.5,43.5){$d$}
\put(-9.5,-3.5){$a$}
\put(44.5,-3.5){$b$}
\put(44.5,43.5){$c$}
\put(8,47.5){$p$}
\put(28,47.5){$p'$}
\put(-13,9){$q$}
\put(-13,29){$q'$}
\end{picture}
\setlength{\unitlength}{0.43mm}
\hskip30pt
\begin{picture}(40,50)
\put(0,0){\line(1,0){45}}
\put(0,0){\line(0,1){45}}
\put(45,0){\line(0,1){45}}
\put(0,45){\line(1,0){45}}
\put(0,0){\circle{2}}
\put(45,0){\circle{2}}
\put(45,45){\circle{2}}
\put(0,45){\circle{2}}
\put(0,50){$a$}
\put(0,-8){$a$}
\put(45,-8){$b$}
\put(45,50){$b$}
\put(20,39){{$n$}}
\put(20,2){{$n$}}
\put(-7,20){$0$}
\put(47,20){$0$}
\end{picture}\hskip35pt
\begin{picture}(40,50)
\put(0,0){\line(1,0){45}}
\put(0,0){\line(0,1){45}}
\put(45,0){\line(0,1){45}}
\put(0,45){\line(1,0){45}}
\put(0,0){\circle{2}}
\put(45,0){\circle{2}}
\put(45,45){\circle{2}}
\put(0,45){\circle{2}}
\put(0,50){$a$}
\put(0,-8){$a$}
\put(45,-8){$b$}
\put(35,50){$b-1$}
\put(12,38){{$n+1$}}
\put(20,2){{$n$}}
\put(-7,20){$0$}
\put(47,20){$1$}
\end{picture}\hskip40pt
\begin{picture}(40,50)
\put(0,0){\line(1,0){45}}
\put(0,0){\line(0,1){45}}
\put(45,0){\line(0,1){45}}
\put(0,45){\line(1,0){45}}
\put(0,0){\circle{2}}
\put(45,0){\circle{2}}
\put(45,45){\circle{2}}
\put(0,45){\circle{2}}
\put(-5,50){$a-1$}
\put(0,-8){$a$}
\put(45,-8){$b$}
\put(45,50){$b$}
\put(12,38){{$n-1$}}
\put(20,2){{$n$}}
\put(-7,20){$1$}
\put(47,20){$0$}
\end{picture}\hskip35pt
\begin{picture}(40,50)
\put(0,0){\line(1,0){45}}
\put(0,0){\line(0,1){45}}
\put(45,0){\line(0,1){45}}
\put(0,45){\line(1,0){45}}
\put(0,0){\circle{2}}
\put(45,0){\circle{2}}
\put(45,45){\circle{2}}
\put(0,45){\circle{2}}
\put(-5,50){$a-1$}
\put(0,-8){$a$}
\put(45,-8){$b$}
\put(35,50){$b-1$}
\put(20,39){{$n$}}
\put(20,2){{$n$}}
\put(-7,20){$1$}
\put(47,20){$1$}
\end{picture}
\caption{The star weight and the four nonvanishing
square weights for $\tau_2(t_q)$}
\end{center}
\label{fig1}
\end{figure}
We consider as in \cite{BBP} a star consisting of four chiral Potts
weights, shown in Fig.~1,
\begin{eqnarray}
U_{p'pq'q}(a,b,c,d)\equiv\sum_{e=1}^N
W_{pq}(a-e)\wb_{pq'}(e-d)\wb_{p'q}(b-e)W_{p'q'}(e-c).
\label{square}
\end{eqnarray}
For the case $\{x_{q'},y_{q'},\mu_{q'}\}\!=
\!\{y\vp_q,\omega^2 x\vp_q,\mu_q^{-1}\}$, it was shown in \cite{BBP}
that 
\begin{equation}
\fl U_{p'pq'q}(a,b,c,d)=0\quad\hbox{for }
0\le \alpha\le1\hbox{ and }2\le \beta\le N-1;
\quad\alpha\equiv a-d,\; \beta\equiv b-c.
\label{square1}
\end{equation}
The product of two transfer matrices becomes a direct sum
of $\tau_2(t_q)$ and $\tau_{N-2}(t_q)$, where the four nonvanishing
configurations of $\tau_2(t_q)$ are shown in Fig.~1. We have
$U_{p'pq'q}(a,b,c,d)\to C_{p'pq}U_{p'pq}^{(2)}(a,b,c,d)$,
with $C_{p'pq}$ some constant given in \cite{BBP}, and
\begin{equation}
\fl U_{p'pq}^{(2)}(a,b,c,d)=\mu_p^{\alpha}\mu_{p'}^{\beta}
\left[\biggl(\frac1{y_p}\biggr)^{\alpha}
\biggl(-\frac{\omega t_q}{y_{p'}}\biggr)^{\beta}
 +\omega^{d-b}\biggl(-\frac{\omega t_q}{y_p y_{p'}}\biggr)
\biggl(\frac{x_p}{t_q}\biggr)^{\alpha}
\biggl(-\omega x_{p'}\biggr)^{\beta}\right],
\label{square2}
\end{equation}
which is related to equation (14) of Baxter \cite{Baxter-tau} by
\begin{equation}
W_{\tau}(t_q|a,b,c,d)=(-\omega t_q)^{a-d-b+c}\,U_{p'pq}^{(2)}(a,b,c,d).
\label{compare1}
\end{equation}
The factor in front cancels out upon multiplying adjacent squares
together, leaving $\tau_2(t_q)$ the same.
Replacing $p,p'$ by $r,r'$ in (\ref{square2}), and letting
$\{x_{r'},y_{r'},\mu_{r'}\}\!=\!\{y\vp_r,\omega^2x\vp_r,\mu_r^{-1}\}$
we find that the square is nonzero for $0\le d-c,a-b\le 1$, and the
nonzero elements in (\ref{square2}) become proportional to weights of
a six-vertex model, namely
\begin{equation}
\fl \biggl(\frac{\mu_r}{y_r}\biggr)^{\!\beta-\!\alpha}
U^{(2)}_{r'rq}(a,b,c,d)\to U_{rq}^{(2,2)} (a,b,c,d)\!=
\biggl(-\frac{t_q}{\omega t_{r}}\biggr)^{\!\beta}
\!-(-1)^{\!\beta}\omega^{d-c-1}
\biggl(\frac{t_q}{t_{r}}\biggr)^{\!1\!-\!\alpha},
\label{6-vertex}
\end{equation}
which is related to equation (5) of Baxter in \cite{Baxter-tau} by%
\footnote{One sign in the third member of (5) in \cite{Baxter-tau} is
misprinted; see also the third item in Figure 2 there.}
\begin{equation}
W_{6v}(t_r/t_q|a,b,c,d)=
(\omega t_r/t_q)( t_q/t_r)^{b-a-c+d}\,U_{rq}^{(2,2)}(b,c,d,a),
\label{compare2}
\end{equation}
in which the vertices are cyclicly permuted.

Consequently, the Yang--Baxter equation of the chiral Potts model
becomes the Yang--Baxter equation for these squares
\begin{eqnarray}
\fl\sum^{N}_{g=1}{U}_{p'pr}^{(2)}(a,g,e,f)
{U}_{p'pq}^{(2)}(b,c,g,a){U}_{rq}^{(2,2)}(c,d,e,g)
\nonumber\\
=\sum^{N}_{g=1}U_{rq}^{(2,2)}(b,g,f,a)
{U}_{p'pq}^{(2)}(g,d,e,f){U}_{p'pr}^{(2)}(b,c,d,g),
\label{YBE}
\end{eqnarray}
which is equation (17) of Baxter \cite{Baxter-tau}.
The product of $L$ such squares, ${\bf U}(t_q)$, has trace
$\tau_2(t_q)$ when the cyclic boundary condition
$\sigma_{L+1}=\sigma_1$ and $\sigma'_{L+1}=\sigma'_1$ is imposed, i.e.
\begin{equation}
\tau_2(t_q)=\prod_{J=1}^L U_{p'pq}^{(2)}
(\sigma\vp_J,\sigma\vp_{J+1},\sigma'_{J+1},\sigma'_J)=\tr{\bf U}(t_q).
\label{tau2}
\end{equation}
To make this more precise, we can go from the Interaction-Round-a-Face
language to te vertex-model language writing
\begin{equation}
\fl n\vp_J=\sigma\vp_J-\sigma\vp_{J+1},\quad
n'_J=\sigma'_J-\sigma'_{J+1},\quad
\alpha\vp_J=\sigma\vp_J-\sigma'_J,\quad
\beta\vp_J=\sigma\vp_{J+1}-\sigma'_{J+1},
\end{equation}
with subtraction mod $N$. Then we define the $2\times2$ monodromy matrix
whose elements are $N^L\times N^L$ matrix functions of $t_q$, i.e.
\begin{equation}
\prod_{J=1}^L U_{p'pq}^{(2)}
(\sigma_J,\sigma_{J+1},\sigma'_{J+1},\sigma'_J)=
{\bf U}(t_q;\{n\vp_J\},\{n'_J\})_{\alpha_1,\beta_L}.
\end{equation}
If we take the $2\times2$ trace implying $\beta_L=\alpha_1$, we must
have $\sigma\vp_{L+1}-\sigma\vp_1=\sigma'_{L+1}-\sigma'_1=m$. Thus
we find $N$ disjoint sectors with boundary condition given by a fixed
jump $m$ mod $N$, $\sigma_{L+1}=\sigma_1+m$, across the boundary.
The sector $m=0$ corresponds to periodic boundary conditions.

Following common practice we write
\begin{equation}
{\bf U}(t_q)=
\left[\begin{array}{cc}
{\bf A}(t_q) &{\bf B}(t_q)\\
{\bf C}(t_q) &{\bf D}(t_q)\end{array}\right]=
\sum_{j=0}^L\,(-\omega t)^j
\left[\begin{array}{cc}
{\bf A}_j &{\bf B}_j\\
{\bf C}_j &{\bf D}_j\end{array}\right],\quad t=t_q/c_{pp'},
\label{ABCD}
\end{equation}
where $c_{pp'}$ is some constant. This satisfies a Yang--Baxter
equation like (\ref{YBE}). Since the $U_{rq}^{(2,2)}$ are the weights
of a six-vertex model, ${\bf U}(t_q)$ intertwines a spin $\frac12$ and
a cyclic representation of quantum group
$\mathrm{U}_q(\widehat{\mathfrak{sl}}_2)$ \cite{JimboNK}. This structure
is intimately related to that on the XXZ model \cite{DFM,Degu3}.

From the Yang--Baxter equation (\ref{YBE}), we find sixteen relations
between the four elements of ${\bf U}(t_q)$ in (\ref{ABCD}), eight of
which are the four three-term relations
\begin{eqnarray}
\fl(1-\omega^{-1}x /y){\bf A}(x){\bf B}(y)=
(1-x /y){\bf B}(y){\bf A}(x)+(1-\omega^{-1}){\bf A}(y){\bf B}(x),
\label{YBE1}\\
\fl(1-\omega^{-1}x /y){\bf A}(y){\bf C}(x)=
\omega^{-1}(1-x /y){\bf C}(x){\bf A}(y)+
(1-\omega^{-1}){\bf A}(x){\bf C}(y),
\label{YBE2}\\
\fl(1-\omega^{-1}x /y){\bf C}(x){\bf D}(y)=
(1-x /y){\bf D}(y){\bf C}(x)+(1-\omega^{-1}){\bf C}(y){\bf D}(x),
\label{YBE3}\\
\fl(1-\omega^{-1}x /y){\bf B}(y){\bf D}(x)=
\omega^{-1}(1-x /y){\bf D}(x){\bf B}(y)+
(1-\omega^{-1}){\bf B}(x){\bf D}(y),
\label{YBE4}
\end{eqnarray}
together with the four commutator relations
\begin{equation}
[{\bf A}(x),{\bf A}(y)]=[{\bf B}(x),{\bf B}(y)]=
[{\bf C}(x),{\bf C}(y)]=[{\bf D}(x),{\bf D}(y)]=0,
\label{fourcomm}
\end{equation}
where $x=t_q$ and $y=t_r$. We shall not use the other eight relations.

%%%%%%%%%%%%%%%%%%%%%%%%%%%%%%%%%%%%%%%%%%%%%%%%%%%%%%%%%%%%%%%%%%%%%%%%
\subsection{Superintegrable $\tau_2(t_q)$}

Now we restrict ourselves to the superintegrable case with
$\{x_{p'},y_{p'},\mu_{p'}\}=\{y_p,x_p,1/\mu_p\}$. After dropping
the subscripts and the factors $(\mu_p/y_p)^{\alpha-\beta}$, which
can be done only for the homogeneous case, the nonvanishing squares
in (\ref{square2}) are
\begin{eqnarray}
U^{(2)}(a,b,b,a)=1-\omega^{a-b+1}t
&&\to {\bf 1}-\omega t{\bf Z},\nonumber\\
U^{(2)}(a,b,b\!-\!1,a)\!=\!-\omega t (1\!-\!\omega^{a-b+1})
&& \to -\omega t ({\bf 1}-{\bf Z}){\bf X}\equiv
-\omega t(1-\omega)\mbox{\bfrak f},\nonumber\\
U^{(2)}(a,b,b,a-1)=1-\omega^{a-b}
&&\to{\bf X}^{-1}({\bf 1}-{\bf Z})\equiv
(1-\omega)\mbox{\bfrak e},\nonumber\\
U^{(2)}(a,b,b\!-\!1,a\!-\!1)\!=\!\omega(\omega^{a-b\!}-\!t)
&&\to\omega{\bf Z}-\omega t{\bf 1},
\label{U2}
\end{eqnarray}
where $t=t_q/t_p$, or $c_{pp'}=t_p$ in (\ref{ABCD}). As these squares
are functions of the differences of the pairs of adjacent spins,
defined in \cite{APsu1} as the edge variables $n=a-b$, we have defined
operators acting on the edge variables given by
\begin{equation}
\fl{\bf Z}_{m,n}=\delta_{m,n}\omega^n,\;
{\bf Z}|n\rangle=\omega^n|n\rangle,\quad
{\bf X}_{m,n}=\delta_{m,n+1},\;
{\bf X}|n\rangle=|n+1\rangle,
\quad n=a-b.
\label{ZX}
\end{equation}
This can be extended to $L$ edges $n_j=\sigma_j-\sigma_{j+1}$ for
$j=1,\cdots,L$, as
\begin{equation} 
\fl{\bf X}_j=\hbox{\vbox{\hbox{$_1$}\hbox{${\bf 1}$}}}\otimes
\hbox{\vbox{\hbox{\;$_{\cdots}$}\hbox{$\cdots$}}}\otimes{\bf 1}\otimes
\hbox{\vbox{\hbox{\;$_j$}\hbox{${\bf X}$}}}\otimes
{\bf 1}\otimes\hbox{\vbox{\hbox{\;$_{\cdots}$}\hbox{$\cdots$}}}
\otimes\hbox{\vbox{\hbox{$_L$}\hbox{${\bf 1}$}}},\qquad
{\bf Z}_j=\hbox{\vbox{\hbox{$_1$}\hbox{${\bf 1}$}}}\otimes
\hbox{\vbox{\hbox{\;$_{\cdots}$}\hbox{$\cdots$}}}\otimes{\bf 1}
\otimes\hbox{\vbox{\hbox{\;$_j$}\hbox{${\bf Z}$}}}\otimes
{\bf 1}\otimes\hbox{\vbox{\hbox{\;$_{\cdots}$}\hbox{$\cdots$}}}
\otimes\hbox{\vbox{\hbox{$_L$}\hbox{${\bf 1}$}}}.
\label{XZj}
\end{equation}
The periodic boundary condition is equivalent to
$n_1+\cdots+n_L\equiv 0\;(\hbox{mod}\,N)$; thus there are only
$N^{L-1}$ independent edge variables. As the products of the squares
${\bf U}$ in (\ref{ABCD}) are functions of the edges variables
only, the transfer matrix $\tau_2(t_q)$---being the trace over
the $N^L$ spin states---is block cyclic. Each block has size
$N^{L-1}\times N^{L-1}$ and $\tau_2(t_q)$ becomes block-diagonal
after Fourier transform, with the $N$ diagonal blocks
\begin{equation}
\tau_2(t_q)|_Q=
{\bf A}(t_q)+\omega^{-Q}{\bf D}(t_q),\qquad Q=0,\cdots, N-1.
\label{tau2q}
\end{equation}
The leading coefficients in (\ref{ABCD}) are easily found,
see (I.25) and (I.26)\footnote{All equations in~\cite{APsu1} are
denoted here by prefacing I to the equation number, those
in~\cite{APsu2} by prefacing II, and those in~\cite{APsu3} by
adding III.},
\begin{eqnarray}
\fl{\bf A}_0={\bf D}_L={\bf 1},\qquad
{\bf A}_L={\bf D}_0\,\omega^{-L}=\prod_{j=1}^L\bZ_j,
\qquad{\bf C}_L={\bf B}_0=0,
\label{AD}\\
\fl{\bf B}_L=(1-\omega)
\sum_{j=1}^L\prod_{m=1}^{j-1}\bZ_m{\mbox{\bfrak f}}_j,\qquad
{\bf C}_0=(1-\omega)
\sum_{j=1}^L\omega^{j-1}\prod_{m=1}^{j-1}\bZ_m{\mbox{\bfrak e}}_j,
\nonumber\\ 
\fl{\bf B}_1=(1-\omega)
\sum_{j=1}^L\,\omega^{L-j}{\mbox{\bfrak f}}_j\prod_{m=j+1}^{L}\bZ_m,
\quad
{\bf C}_{L-1}=(1-\omega)
\sum_{j=1}^L\,{\mbox{\bfrak e}}_j\prod_{m=j+1}^{L}\bZ_m.
\label{BC}
\end{eqnarray} 

%%%%%%%%%%%%%%%%%%%%%%%%%%%%%%%%%%%%%%%%%%%%%%%%%%%%%%%%%%%%%%%%%%%%%%%%
\subsection{Relationship with generators of $U_q({\mathfrak{sl}}_2)$}

The generators $\mbox{\bfrak e}_j$ and $\mbox{\bfrak f}_j$ in the
above equations are defined by
\begin{equation} 
(1-\omega)\mbox{\bfrak e}_j={\bf X}_j^{-1}({\bf 1}-{\bf Z}_j),\quad
(1-\omega)\mbox{\bfrak f}_j=({\bf 1}-{\bf Z}_j){\bf X}_j,
\label{ef}
\end{equation}
and satisfy the relation
\begin{equation} 
(1-\omega)(\mbox{\bfrak e}\vp_j\mbox{\bfrak f}\vp_j-
\omega\mbox{\bfrak f}\vp_j\mbox{\bfrak e}\vp_j)
={\bf 1}-\omega{\bf Z}_j^2.
\end{equation}
They are not the same as the usual $\mbox{\bfrak e}'_j$ and
$\mbox{\bfrak f}'_j$ of the quantum group $U_q({\mathfrak{sl}}_2)$,
but are related by
\begin{equation}
\mbox{\bfrak e}'_j=-q\mbox{\bfrak e}\vp_j{\bf Z}_j^{-1/2},\quad
\mbox{\bfrak f}'_j=q{\bf Z}_j^{-1/2}\mbox{\bfrak f}_j,\quad
\omega=q^2,
\end{equation}
compare the equation below (4.4) in \cite{NiDe2}. Operators
$\mbox{\bfrak e}'_j$ and $\mbox{\bfrak f}'_j$ satisfy the relation
\begin{equation} 
(q-q^{-1})(\mbox{\bfrak e}'_j\mbox{\bfrak f}'_j-
\mbox{\bfrak f}'_j\mbox{\bfrak e}'_j)
=q{\bf Z}_j-(q{\bf Z}_j)^{-1},
\end{equation}
as defined by Jimbo \cite{JimboNK}. This difference in these operators 
is due to the fact that the six-vertex model in (\ref{6-vertex}) is
not symmetric.

%%%%%%%%%%%%%%%%%%%%%%%%%%%%%%%%%%%%%%%%%%%%%%%%%%%%%%%%%%%%%%%%%%%%%%%%
\subsection{Commutation relations}

We use (\ref{YBE1}) to (\ref{YBE4}) to derive commutation
relations. Equating the coefficients of $x^{L+1}$ in (\ref{YBE1})
and the coefficients of $x^{0}$ in (\ref{YBE4}) where ${\bf B}_0=0$,
we find
\begin{equation}
{\bf A}_L{\bf B}(y)=\omega{\bf B}(y){\bf A}_L,\quad
{\bf D}_0{\bf B}(y)=\omega{\bf B}(y){\bf D}_0.
\label{com1}
\end{equation}
In the limit $y\to 0$, we have ${\bf B}(y)\to-\omega y{\bf B}_1$
as ${\bf B}_0=0$, so that (\ref{YBE1}) becomes
\begin{equation}
{\bf A}(x){\bf B}_1 -\omega{\bf B}_1{\bf A}(x)=
(1-\omega^{-1})x^{-1}{\bf A}_0{\bf B}(x)
=(1-\omega^{-1})x^{-1}{\bf B}(x),
\label{com2}
\end{equation}
using ${\bf A}_0={\bf 1}$. By equating the coefficients of $y^{L}$
in (\ref{YBE1}), we find
\begin{equation}
{\bf A}(x){\bf B}_L-{\bf B}_L{\bf A}(x)=
(1-\omega^{-1}){\bf A}_L{\bf B}(x)
=(\omega-1){\bf B}(x){\bf A}_L,
\label{com3}
\end{equation}
where (\ref{com1}) has been used. Similarly, equating the coefficients
of $y^L$ in (\ref{YBE2}) and of $y^{-1}$ in (\ref{YBE3}), and also the
coefficients of $x^0$ and $x^L$ in (\ref{YBE2}), we find
\begin{eqnarray}
{\bf A}_L{\bf C}(x) =\omega^{-1}{\bf C}(x){\bf A}_L,\qquad
{\bf D}_0{\bf C}(x) =\omega^{-1}{\bf C}(x){\bf D}_0,\\
{\bf A}(y){\bf C}_0 -\omega^{-1}{\bf C}_0{\bf A}(y)=
(1-\omega^{-1}){\bf C}(y),
\nonumber\\
{\bf A}(y){\bf C}_{L-1}-{\bf C}_{L-1}{\bf A}(y)
=(\omega-1)y{\bf C}(y){\bf A}_L,
\label{com4}
\end{eqnarray}
using ${\bf C}_L=0$ and ${\bf A}_0={\bf 1}$. In the same way,
(\ref{YBE3}) and (\ref{YBE4}) yield the relations
\begin{eqnarray}
{\bf D}(y){\bf C}_0 -{\bf C}_0{\bf D}(y)=
-(1-\omega^{-1}){\bf C}(y){\bf D}_0,
\nonumber\\
{\bf D}(y){\bf C}_{L-1}-\omega^{-1}{\bf C}_{L-1}{\bf D}(x)
=-(\omega-1)y{\bf C}(y),
\nonumber\\
{\bf D}(x){\bf B}_1-{\bf B}_1{\bf D}(x)
=-(1-\omega^{-1})x^{-1}{\bf B}(x){\bf D}_0,
\nonumber\\
{\bf D}(x){\bf B}_L-\omega{\bf B}_L{\bf D}(x)
=-(\omega-1){\bf B}(x).
\label{com5}
\end{eqnarray}

Using (\ref{com1}) through (\ref{com5}) and (\ref{fourcomm})
it is straightforward to prove by induction the following relations,
\begin{eqnarray}
{\bf A}(x){\bf C}^n_0=\omega^{-n}{\bf C}^n_0{\bf A}(x)+
(\omega-1)\omega^{-n}[n]{\bf C}^{n-1}_0{\bf C}(x),
\label{ind1}\\
{\bf D}(x){\bf C}^n_0={\bf C}^n_0{\bf D}(x)
-(\omega-1)\omega^{-n}[n]{\bf C}^{n-1}_0{\bf C}(x){\bf D}_0,
\label{ind2}\\
{\bf A}(x){\bf B}^n_1=\omega^n{\bf B}^n_1{\bf A}(x)+
(1-\omega^{-1})x^{-1}[n]{\bf B}^{n-1}_1{\bf B}(x),
\label{ind3}\\ 
{\bf D}(x){\bf B}^n_1={\bf B}^n_1{\bf D}(x)-
(1-\omega^{-1})x^{-1}[n]{\bf B}^{n-1}_1{\bf B}(x){\bf D}_0,
\label{ind4}
\end{eqnarray}
where $[n]\equiv1+\cdots +\omega^{n-1}$. Similar relations
for ${\bf B}_{L}$ and ${\bf C}_{L-1}$ are
\begin{eqnarray}
{\bf A}(x){\bf C}^n_{L-1}={\bf C}^n_{L-1}{\bf A}(x)+
(\omega-1)\omega^{1-n}x[n]{\bf C}^{n-1}_{L-1}{\bf C}(x){\bf A}\vp_L,
\label{ind5}\\
{\bf D}(x){\bf C}^n_{L-1}=\omega^{-n}{\bf C}^n_{L-1}
{\bf D}(x)-(\omega-1)\omega^{1-n}x[n]{\bf C}^{n-1}_{L-1}{\bf C}(x),
\label{ind6}\\
{\bf A}(x){\bf B}^n_L={\bf B}^n_L
{\bf A}(x)+(\omega-1)[n]{\bf B}^{n-1}_L{\bf B}(x){\bf A}\vp_L,
\label{ind7}\\ 
{\bf D}(x){\bf B}^n_L=\omega^n{\bf B}^n_L{\bf D}(x)-
(\omega-1)[n]{\bf B}^{n-1}_L{\bf B}(x).
\label{ind8}
\end{eqnarray}

%%%%%%%%%%%%%%%%%%%%%%%%%%%%%%%%%%%%%%%%%%%%%%%%%%%%%%%%%%%%%%%%%%%%%%%%
\section{Eigenvectors of $\tau_2(t_q)|_Q$}

We shall find the eigenvectors $ \nu_Q$ of $\tau_2(t_q)|_Q$ such that
\begin{eqnarray}
\tau_2(t_q)|\vp_Q\,\nu_Q=
[(1-\omega t)^L+\omega^{-Q}(1-t)^L]\nu_Q,\quad\hbox{or}
\label{eigv1}\\
\tau_2(t_q)|\vp_Q\,\nu_Q=
[\omega^{-Q}(1-\omega t)^L+(1-t)^L]\nu_Q,
\label{eigv2}
\end{eqnarray}
where $t=t_q/t_p$. Defining similarly as in \cite{NiDe2}
\begin{equation}
B_j^{(n)}=\lim_{\textstyle{{q\to\omega}\atop{|q|<1}}}
\frac{B_j^n}{[n]!},\quad\hbox{with }
[n]=\frac{1-q^n}{1-q},\quad[n]!=[n]\cdots[2]\,[1],
\label{facto}
\end{equation}
and using (\ref{ind1}) and (\ref{ind3}), we can show
\begin{eqnarray}
\fl{\bf A}(x){\bf C}_0^{(nN+Q)}{\bf B}_1^{(mN+Q)}
\nonumber\\
=\omega^{-Q}\Big[{\bf C}_0^{(nN+Q)}{\bf A}(x)
+(\omega-1){\bf C}^{(nN+Q-1)}_0{\bf C}(x)\Big]{\bf B}_1^{(mN+Q)}
\nonumber\\
=\omega^{-Q}{\bf C}_0^{(nN+Q)}\Big[\omega^Q{\bf B}^{(mN+Q)}_1{\bf A}(x)
+(1-\omega^{-1})x^{-1}{\bf B}^{(mN+Q-1)}_1{\bf B}(x)\Big]
\nonumber\\
\quad+\omega^{-Q}(\omega-1){\bf C}^{(nN+Q-1)}_0
{\bf C}(x){\bf B}^{(mN+Q)}_1,
\label{ACBQ}
\end{eqnarray}
while (\ref{ind2}), (\ref{ind4}) and (\ref{com1}) yield
\begin{eqnarray}
\fl{\bf D}(x){\bf C}_0^{(nN+Q)}{\bf B}_1^{(mN+Q)}
\nonumber\\
={\bf C}_0^{(nN+Q)}\Big[{\bf B}^{(mN+Q)}_1{\bf D}(x)-
(1-\omega^{-1})x^{-1}{\bf B}^{(mN+Q-1)}_1{\bf B}(x)\Big]{\bf D}\vp_0
\nonumber\\
\quad-(\omega-1){\bf C}^{(nN+Q-1)}_0{\bf C}(x)
{\bf B}^{(mN+Q)}_1{\bf D}\vp_0.
\label{DCBQ}
\end{eqnarray}
Consequently, we find that
\begin{eqnarray}
\fl\Big[{\bf A}(x)+\omega^{-Q}{\bf D}(x),
{\bf C}_0^{(nN+Q)}{\bf B}_1^{(mN+Q)}\Big]
\nonumber\\
=\omega^{-Q}(\omega-1)\Big[(\omega x)^{-1}{\bf C}_0^{(nN+Q)}
{\bf B}^{(mN+Q-1)}_1{\bf B}(x)
\nonumber\\
\quad+{\bf C}^{(nN+Q-1)}_0{\bf C}(x){\bf B}^{(mN+Q)}_1\Big]
({\bf 1}-{\bf D}\vp_0).
\label{tauCBq}
\end{eqnarray}
From (\ref{AD}) we have
${\bf D}_0=\omega^{L}\prod_{j=1}^L\bZ_j$, so that for $L$ a multiple
of $N$, and for $|\{n_j\}\rangle$ with
$n_1+\cdots+n_L\equiv0\;(\mathrm{mod}\,N)$, we have
$({\bf 1}-{\bf D}_0)|\{n_j\}\rangle=0$. Hence,
\begin{equation}
\Big[{\bf A}(x)+\omega^{-Q}{\bf D}(x),
{\bf C}_{0}^{(nN+Q)}{\bf B}_1^{(mN+Q)}\Big]\,|\{n_j\}\rangle=0.
\label{tauCB}
\end{equation} 
Similarly, we can prove
\begin{eqnarray}
&&\Big[{\bf A}(x)+\;\omega^{Q}\,{\bf D}(x),
{\bf B}_1^{(mN+Q)}{\bf C}_0^{(nN+Q)}\Big]\,|\{n_j\}\rangle=0,
\nonumber\\
&&\Big[{\bf A}(x)+\omega^{-Q}{\bf D}(x),
{\bf B}_L^{(mN+Q)}{\bf C}_{L-1}^{(nN+Q)}\Big]\,|\{n_j\}\rangle=0,
\nonumber\\
&&\Big[{\bf A}(x)+\;\omega^{Q}\,{\bf D}(x),
{\bf C}_{L-1}^{(nN+Q)}{\bf B}_L^{(mN+Q)}\Big]\,|\{n_j\}\rangle=0.
\label{tauBC1L}
\end{eqnarray}
Particularly, the ferromagnetic ground state
$|\Omega\rangle\!=\!|\{0\}\rangle$ and the antiferromagnetic
ground state $|{\bar\Omega}\rangle\!=\!|\{N-1\}\rangle$
are easily seen, from either (\ref{U2}) or (I.29), to satisfy
\begin{eqnarray}
\tau_2(t_q)|\vp_Q\,|\Omega\rangle=
[(1-\omega t)^L\!+\omega^{-Q}(1-t)^L]\,|\Omega\rangle,
\label{ground1}\\
\tau_2(t_q)|\vp_Q\,|{\bar\Omega}\rangle=
[\omega^{-Q}(1-\omega t)^L\!+(1-t)^L]\,|{\bar\Omega}\rangle.
\label{ground2}
\end{eqnarray}
Due to (\ref{tauCB}) and (\ref{tauBC1L}), we find that
\begin{equation}
\prod_{j=1}^J{\bf C}_0^{(m_jN+Q)}{\bf B}_1^{(n_jN+Q)}|\Omega\rangle,
\quad\prod_{j=1}^J{\bf C}_{L-1}^{(m_jN+N-Q)}
{\bf B}_L^{(n_jN+N-Q)}|\Omega\rangle
\label{eigvec1}
\end{equation}
are eigenvectors in the same degenerate eigenspace as
$|\Omega\rangle$, while
\begin{equation}
\prod_{j=1}^J{\bf B}_1^{(m_jN+N-Q)}
{\bf C}_0^{(n_j N+N-Q)}|{\bar\Omega}\rangle,
\quad\prod_{j=1}^J{\bf B}_L^{(m_jN+Q)}
{\bf C}_{L-1}^{(n_jN+Q)}|{\bar\Omega}\rangle
\label{eigvec2}
\end{equation}
are eigenvectors in the same degenerate eigenspace as
$|{\bar\Omega}\rangle$. For $Q\ne0$, we conclude from calculations
for $N,L$ small, that these two eigenspaces have dimension $2^{r-1}$
($m_Q=r-1$). Thus by letting $0\le m_1<n_1<\cdots<m_J<n_J\le r-1$,
where $0\le J\le r-1$ and $\sum_{j=1}^J (n_j-m_j)=J$, similar to the
results given in \cite{ItoTer}, we can obtain a basis of $2^{r-1}$
eigenvectors in each of the two eigenspaces corresponding to the two
eigenvalues. For $Q=0$, it is easily seen from (\ref{ground1}) and 
(\ref{ground2}) that the two eigenvalues become equal and the two
eigenspaces merge into one.

From (I.47) in \cite{APsu1}, we find an other way to obtain $\tau_2$
eigenvectors of the two degenerate eigenspaces, but this leads to
the complication that one has to deal with the $\pm Q$ sectors at the
same time. It is far from obvious how to find the generators of the
loop algebra. We now use the information obtained in the $Q=0$ case
\cite{APsu2} to find what we believe to be the best choices
in the $Q\ne 0$ cases.

%%%%%%%%%%%%%%%%%%%%%%%%%%%%%%%%%%%%%%%%%%%%%%%%%%%%%%%%%%%%%%%%%%%%%%%%
\section{Quantum Loop Subalgebra}

We now shall present the generators of the ${\mathfrak{sl}}_2$ algebras
and of the loop (sub)algebras for $Q\ne 0$. Since the eigenvalues of
the transfer matrices are Ising-like, and the eigenspaces for $\tau_2$
are highly degenerate for $Q\ne0$ as well, we want to construct the loop
algebra on the eigenvectors of this degenerate eigenspace, in the same
way as is done for the $Q=0$ case in \cite{NiDe2,APsu2}. Once this is
accomplished, we can obtain the generators of the ${\mathfrak{sl}}_2$
algebras. In subsection \ref{drin}, we first generalize (I.39)
to obtain operators on the ground state, which yield the coefficients
of the Drinfeld polynomials $P_Q(z)$. Then, in subsection \ref{Epm},
we generalize (II.52) through (II.54) to obtain the expressions for
${\bf E}^{\pm}_{m,Q}$ on the ground state. In order to define the
action of ${\bf E}^{\pm}_{m,Q}$ on other states, we have to
generalize the construction for the case $Q=0$ in \cite{APsu2},
where we defined ${\bf E}^{\pm}_{m,0}$ in terms of loop algebra
generators ${\bf x}^{\pm}_{m,0}$. To do so,
in subsection \ref{xpm}, we generalize identities (II.50), (II.12),
(II.45) through (II.47), and obtain the expressions for the generators
${\bf x}^{\pm}_{m,Q}$ acting on the ground state. In subsection
\ref{hm}, we show that the necessary condition that these operators
generate a loop algebra, or subalgebra, is satisfied. This enables
us to propose the generators of the loop subalgebra in subsection
\ref{loop}. The reader may skip the remainder of this section.
 
%%%%%%%%%%%%%%%%%%%%%%%%%%%%%%%%%%%%%%%%%%%%%%%%%%%%%%%%%%%%%%%%%%%%%%%%
\subsection{Drinfeld polynomials\label{drin}}

From (\ref{BC}) and the identities ${\bf Z}\vp_i{\mbox{\bfrak f}}_j=
\omega\delta\vp_{ij}{\mbox{\bfrak f}}_j{\bf Z}\vp_i$,
${\bf Z}\vp_i{\mbox{\bfrak e}}_j=
\omega^{-1}\delta\vp_{ij}{\mbox{\bfrak e}}_j{\bf Z}\vp_i$, we find that
\begin{eqnarray}
\fl{\bfC}_0^{(m)}\equiv{\bf C}_0^{(m)}(1-\omega)^{-m}=
\sum_{{\{0\le n_j\le N-1\}}\atop{n_1+\cdots+n_L=m}}
\prod_{j=1}^{L}
\bZ_j^{\bN_{j}}\, \frac{\omega_{\vp}^{(j-1)n_j}
{\mbox{\bfrak e}}_j^{n_j}}{[n_j]!},\quad
\bN_j=\sum_{\ell=j+1}^L n_\ell,\nonumber\\
\fl{\bfB}_1^{(m)}\equiv{\bf B}_1^{(m)}(1-\omega)^{-m}=
\sum_{{\{0\le n_j\le N-1\}}\atop{n_1+\cdots+n_L=m}}
\prod_{j=1}^{L}\frac{\omega_{\vp}^{-j n_j}
{\mbox{\bfrak f}}_j^{n_j}}{[n_j]!}\,
\bZ_j^{N_{j}},\qquad N_j=\sum_{\ell=1}^{j-1}n_\ell.
\label{C0B1m}
\end{eqnarray}
Using (\ref{ef}) or (II.55), we find
\begin{equation}
{\bfC}_0^{(mN+Q)}{\bfB}_1^{(mN+Q)}|\Omega\rangle
=\omega^{-Q}\sum_{{\{0\le n_j\le N-1\}}\atop{n_1+\cdots+n_L=mN+Q}}
|\Omega\rangle=\omega^{-Q}\Lambda^Q_{m}|\Omega\rangle.
\label{coeff1}
\end{equation}
Here the $\Lambda^Q_{m}$ are the coefficients of the Drinfeld
polynomial $P_Q(z)$ in (\ref{roots}). However, (\ref{BC}) also yields
\begin{eqnarray}
{\bfC}_{L-1}^{(m)}\equiv{\bf C}_{L-1}^{(m)}(1-\omega)^{-m}=
\sum_{{\{0\le n_j\le N-1\}}\atop{n_1+\cdots+n_L=m}}
\prod_{j=1}^{L}\bZ_j^{N_{j}}\,
\frac{{\mbox{\bfrak e}}_j^{n_j}}{[n_j]!},\nonumber\\
{\bfB}_L^{(m)}\equiv{\bf B}_L^{(m)}(1-\omega)^{-m}=
\sum_{{\{0\le n_j\le N-1\}}\atop{n_1+\cdots+n_L=m}}
\prod_{j=1}^{L}\frac{{\mbox{\bfrak f}}_j^{n_j}}{[n_j]!}\,
\bZ_j^{\bN_{j}},
\label{CBm}
\end{eqnarray}
so that
\begin{equation}
{\bfC}_{L-1}^{(mN+N-Q)}{\bfB}_L^{(mN+N-Q)}|\Omega\rangle
=\sum_{{\{0\le n_j\le N-1\}}\atop{n_1+\cdots+n_L=mN+N-Q}}
|\Omega\rangle=\Lambda^{N-Q}_{m}|\Omega\rangle.
\label{coeff2}
\end{equation}
Now the $\Lambda^{N-Q}_{m}$ are the coefficients of the polynomial
$P_{N-Q}(z)$, whose roots are the inverses of the roots of $P_Q(z)$.
We have the situation that the two sets of eigenvectors in
(\ref{eigvec1}) have the same eigenvalues, but correspond to different
Drinfeld polynomials. On the other hand, the coefficients of the
Drinfeld polynomial are symmetric ($\Lambda_m=\Lambda_{r-m}$) for $Q=0$,
so that the roots of the polynomial then appear in pairs $z_j$
and $1/z_j$. Since the algebra and the roots of the Drinfeld polynomials
are intimately related \cite{APsu1,APsu2,Degu2}, the corresponding
algebras for $Q\ne0$ cases are different from the algebra for
the $Q=0$ case. We shall explore this next in more detail.

%%%%%%%%%%%%%%%%%%%%%%%%%%%%%%%%%%%%%%%%%%%%%%%%%%%%%%%%%%%%%%%%%%%%%%%%
\subsection{Generators ${\bf E}^{\pm}_{m,Q}$ on the ground state
\label{Epm}}

In (\ref{roots}), we have let $z_{m,Q}$ denote the roots of the
Drinfeld polynomial $P_Q(z)$. Now, as in (II.10) or 
(III.56) \cite{APsu3,Davies}, we define the polynomials
\begin{equation}
f^Q_j(z)=\prod_{\ell\ne j}{\frac{z-z_{\ell,Q}}{z_{j,Q}-z_{\ell,Q}}}=
\sum_{n=0}^{m_Q-1}\beta^Q_{j,n}z^n, \quad
f^Q_j(z_{k,Q})=\delta_{j,k}\,,
\label{beta}
\end{equation}
where $\beta^Q_{j,n}$ are the elements of the inverse of the
Vandermonde matrix, such that
\begin{equation}
\sum_{n=0}^{m_Q-1}\beta^Q_{j,n}z_{k,Q}^n=\delta_{j,k}, \qquad
\sum_{k=1}^{m_Q}z_{k,Q}^n\beta^Q_{k,m}=\delta_{n,m},
\quad\hbox{for } 0\le n\le m_Q-1.
\label{vdm}
\end{equation}
Thus we generalize the previous results to include the cases
for $Q\ne0$. We may also generalize (II.53) and (II.54) to
\begin{eqnarray}
\langle\Omega|{\bf E}_{m,Q}^{-}
=-\omega^{Q}(\beta^Q_{m,0}/\Lambda^Q_{0})
\sum_{\ell=1}^{m_Q} z_{m,Q}^{\ell-1}
\langle\Omega|{\bfC}_0^{(\ell N+Q)}{\bfB}_1^{(\ell N-N+Q)},
\label{ome}\\
{\bf E}_{m,Q}^+|\Omega\rangle
=\omega^{Q}(\beta^Q_{m,0}/\Lambda^Q_{0})
\sum_{\ell=1}^{m_Q} z_{m,Q}^{\ell}
{\bfC}_0^{(\ell N-N+Q)}{\bfB}_1^{(\ell N+Q)}|\Omega\rangle.
\label{epo}
\end{eqnarray}
If the ${\bf E}_{m,Q}^{\pm}$ are to be generators of
${\mathfrak{sl}}_2$ algebras, then it is necessary that
\begin{equation}
\langle\Omega|{\bf E}_{k,Q}^{-}{\bf E}_{m,Q}^+|\Omega\rangle
=-\delta_{k,m}\langle\Omega|{\bf H}_{k}^{Q}|\Omega\rangle
=\delta_{k,m}.
\label{ortho}
\end{equation}
To show this, we use (\ref{C0B1m}) and (II.55) to obtain
\begin{eqnarray}
\fl\langle \Omega|{\bfC}_0^{(\ell N+N+Q)}{\bfB}_1^{(\ell N+Q)}
=\omega^{-Q}\sum_{{\{0\le n_j\le N-1\}}\atop{n_1+\cdots+n_L=N}}
\langle\{n_j\}|\omega^{\sum_{j} j n_j}{\bar K}_{\ell N+Q}(\{n_j\}),
\label{ocbk}\\
\fl{\bfC}_0^{(\ell N+Q)}{\bfB}_1^{(\ell N+N+Q)} |\Omega\rangle
=\omega^{-Q}\sum_{{\{0\le n_j\le N-1\}}\atop{n_1+\cdots+n_L=N}}
\omega^{-\sum_{j}j n_j}K_{\ell N+Q}(\{n_j\})|\{n_j\}\rangle,
\label{cbok}
\end{eqnarray}
where $K_m(\{n_j\})$ and $\bK_m(\{n_j\})$ are defined in (III.7) and
(III.8). Equations (\ref{ocbk}) and (\ref{cbok}) are similar to (II.59).
Substituting them into (\ref{ome}) and (\ref{epo}) with $\ell$
replaced by $\ell+1$, then using (II.63) and (II.64) [or (III.16)],
we find
\begin{eqnarray}
\langle \Omega|{\bf E}_{m,Q}^{-}=-(\beta^Q_{m,0}/\Lambda^Q_{0})
\sum_{{\{0\le n_j\le N-1\}}\atop{n_1+\cdots+n_L=N}}\langle\{n_j\}|\,
\omega^{\sum_{j}j n_j}\bar{G}^{\vp}_Q(\{n_j\},z^{\vp}_{m,Q}),
\label{ome2}\\
{\bf E}_{k,Q}^{+}|\Omega\rangle=(\beta^Q_{k,0}/\Lambda^Q_{0})
z^{\vp}_{k,Q}\sum_{{\{0\le n_j\le N-1\}}\atop{n_1+\cdots+n_L=N}}
\omega^{-\sum_{j}j n_j}
{G}^{\vp}_Q(\{n_j\},z^{\vp}_{k,Q})|\{n_j\}\rangle.
\label{epo2}
\end{eqnarray}
We now use the main theorem in \cite{APsu3} to prove (\ref{ortho}).
From (\ref{beta}), we find
\begin{equation}
\beta^Q_{m,0}=\prod_{\ell\ne m}\frac{-z^{\vp}_{\ell,Q}}
{z^{\vp}_{m,Q}-z^{\vp}_{\ell,Q}}=-\frac{\Lambda^Q_{0}}
{\Lambda^Q_{m_Q}z^{\vp}_{m,Q}}
\prod_{\ell\ne m}\frac1{z^{\vp}_{m,Q}-z^{\vp}_{\ell,Q}},
\label{beta0}
\end{equation}
so that the constant in (III.19) becomes
\begin{equation}
B^{\vp}_{m,Q}=(\Lambda^Q_{m_Q})^2 z^{\vp}_{m,Q}
\prod_{\ell\ne m}(z^{\vp}_{m,Q}-z^{\vp}_{\ell,Q})^2
=(\Lambda^Q_{0}/\beta^Q_{m,0})^2 z^{-1}_{m,Q}.
\label{conB}
\end{equation}
Consequently, we may combine (\ref{ome2}) and (\ref{epo2}), and
then use (III.18) to get
\begin{equation}
\fl\langle\Omega|{\bf E}_{k,Q}^{-}{\bf E}_{m,Q}^+|\Omega\rangle
=-\frac{z^{\vp}_{k,Q}\beta^Q_{k,0}\beta^Q_{m,0}}
{(\Lambda^Q_{0})_{\vp}^2}
\sum_{{\{0\le n_j\le N-1\}}\atop{n_1+\cdots+n_L=N}}
\bar{G}^{\vp}_Q(\{n_j\},z^{\vp}_{m,Q})
{G}^{\vp}_Q(\{n_j\},z^{\vp}_{k,Q})
=\delta_{k,m}.
\label{ortho1}
\end{equation}
This is the first evidence that the above generalization of (II.53)
and (II.54) to $Q\ne0$ cases is correct.

%%%%%%%%%%%%%%%%%%%%%%%%%%%%%%%%%%%%%%%%%%%%%%%%%%%%%%%%%%%%%%%%%%%%%%%%
\subsection{Generators ${\bf x}^{\pm}_{m,Q}$
on the ground state\label{xpm}}

In paper \cite{APsu2}, we have studied the $Q=0$ case, for which the 
generators ${\bf x}^{\pm}_{m}$ of the loop algebra were known from
\cite{APsu1}. From these operators, we obtained the
${\bf E}^{\pm}_{m,0}$, the generators of the ${\mathfrak{sl}}_2$'s.
In this paper, we will go in the  reverse order, by using (\ref{ome})
and (\ref{epo}) to determine the best form of the ${\bf x}^{\pm}_{m,Q}$.
As in (II.50) we let
\begin{equation}
S^Q_n=\sum_{m=1}^{m_Q}\beta^Q_{m,0}z^{-n}_{m,Q}, \qquad S^Q_n=0,\;
\hbox{ for }\; 1-m_Q\le n<0,
\label{sq}
\end{equation}
where the second equation of (\ref{vdm}) has been used to show
$S^Q_n=\delta_{n,0}$ for $1-m_Q\le n\le0$. In fact,
$P_Q(0)/P_Q(z)=\ldots=\sum_{n=0}^{\infty}S_n^Qz^n$
as in (II.49) and using (\ref{beta0}) leads to (\ref{sq}) for
all $n\ge0$.

Now similar to (II.12), we let
\begin{eqnarray}
\fl{\bf x}^{-}_{n,Q}|\Omega\rangle=\sum_{m=1}^{m_Q}
z^{-n}_{m,Q}{\bf E}_{m,Q}^{+}|\Omega\rangle=
\omega^{Q}\sum_{m=1}^{m_Q} z^{-n}_{m,Q}(\beta^Q_{m,0}/\Lambda^Q_{0})
\sum_{\ell=1}^{m_Q} z_{m,Q}^{\ell}
{\bfC}_0^{(\ell N-N+Q)}{\bfB}_1^{(\ell N+Q)}|\Omega\rangle
\nonumber\\
=(\omega^{Q}/\Lambda^Q_{0})\sum_{\ell=1}^{n}S^Q_{n-\ell}
{\bfC}_0^{(\ell N-N+Q)}{\bfB}_1^{(\ell N+Q)}|\Omega\rangle,
\label{xm}
\end{eqnarray}
where (\ref{epo}) and (\ref{sq}) have been used. Similarly we
find from (\ref{ome})
\begin{equation}
\fl\langle\Omega|{\bf x}^{+}_{n,Q}=-\sum_{m=1}^{m_Q} z^{-n}_{m,Q}
\langle\Omega|{\bf E}_{m,Q}^{-}
=(\omega^{Q}/\Lambda^Q_{0})\sum_{\ell=0}^{n}S^Q_{n-\ell}
\langle\Omega|{\bfC}_0^{(\ell N+N+Q)}{\bfB}_1^{(\ell N+Q)}.
\label{xp}
\end{equation}
These are generalizations of (II.45) and (II.46).

Furthermore, the relation (II.47) can be generalized to
\begin{equation}
\sum_{n=0}^m\Lambda^Q_{m-n}S^Q_n=\Lambda^Q_{0}\delta_{m,0}\quad 
\hbox{with }\; S^Q_0=1.
\label{lsq}
\end{equation}
To show this, we note that for $m=0$ we already have $S^Q_0=1$, while
for $m\ge1$ we write
\begin{equation}
\fl\sum_{n=0}^m\Lambda^Q_{m-n}S^Q_n
=\sum_{n=0}^m\Lambda^Q_{n}S^Q_{m-n}
=\sum_{n=0}^{m_Q}\Lambda^Q_{n}\sum_{\ell=1}^{m_Q}
\beta^Q_{\ell,0}z^{n-m}_{\ell,Q}
=\sum_{\ell=1}^{m_Q}\beta^Q_{\ell,0}z^{-m}_{\ell,Q}
\sum_{n=0}^{m_Q}\Lambda^Q_{n}z^{n}_{\ell,Q}=0,
\end{equation}
where the summation over $n$ has been changed to $0\le n\le m_Q$
as $S^Q_{m-n}=0$ for $m<n\le m_Q$, or $\Lambda^Q_n=0$ for
$n>m_Q$. Substituting (\ref{sq}) into the sum and interchanging
the order of summation we find the sum is identically zero for $m>0$,
as the $z_{\ell,Q}$ are roots of the Drinfeld polynomial,
$P_Q(z_{\ell,Q})=0$. Thus (\ref{lsq}) holds for all $m\ge0$.

Using (\ref{xm}) and (\ref{lsq}) we generalize (I.42) to
\begin{eqnarray}
\fl\sum_{n=1}^{m}\Lambda^Q_{m-n}{\bf x}^{-}_{n,Q}|\Omega\rangle
=(\omega^{Q}/\Lambda^Q_{0})\sum_{\ell=1}^{m}
\sum_{n=\ell}^{m}\Lambda^Q_{m-n} S^Q_{n-\ell}
{\bfC}_0^{(\ell N-N+Q)}{\bfB}_1^{(\ell N+Q)}|\Omega\rangle
\nonumber\\
=\omega^{Q}\sum_{\ell=1}^{m}\delta_{m,\ell}
{\bfC}_0^{(\ell N-N+Q)}{\bfB}_1^{(\ell N+Q)}|\Omega\rangle
=\omega^{Q}{\bfC}_0^{(mN-N+Q)}{\bfB}_1^{(m N+Q)}|\Omega\rangle.
\label{invxm}
\end{eqnarray}
For $m>m_Q=\lfloor (L(N-1)+Q)/N\rfloor$, the right hand side is
identically zero, so that there are $m_Q$ independent vectors
${\bf x}^{-}_{n,Q}|\Omega\rangle$.
Similarly, from (\ref{xp}) and (\ref{lsq}) we may derive
\begin{equation}
\sum_{n=0}^{m}\Lambda^Q_{m-n}\langle\Omega|{\bf x}^{+}_{n,Q}
=\omega^{Q}\langle\Omega|{\bfC}_0^{(m N+ N+Q)}{\bfB}_1^{(m N+Q)}.
\label{invxp}
\end{equation}

%%%%%%%%%%%%%%%%%%%%%%%%%%%%%%%%%%%%%%%%%%%%%%%%%%%%%%%%%%%%%%%%%%%%%%%%
\subsection{Generators ${\bf h}_{m,Q}$ on the ground state\label{hm}}

We define
\begin{equation}
d_{m,Q}=\langle\Omega|{\bf h}_{m,Q}|\Omega\rangle=
\langle\Omega|{\bf x}^{+}_{m-1,Q}{\bf x}^{-}_{1,Q}|\Omega\rangle,
\quad\hbox{for}\quad 1\le m<\infty.
\label{dq}
\end{equation}
Substituting (\ref{xm}) and (\ref{xp}) into the above equation and
using (\ref{ocbk}) and (\ref{cbok}), we find
\begin{equation}
d_{m,Q}=(\Lambda^Q_{0})\vf^{-2}\sum_{\ell=0}^{m-1} S^Q_{m-1-\ell}
\sum_{{\{0\le n_j\le N-1\}}\atop{n_1+\cdots+n_L=N}}
\bK^{\vp}_{\ell N+Q}(\{n_j\}){K}^{\vp}_Q(\{n_j\}).
\label{dq1}
\end{equation}
After changing the summation variable $\ell$ by $\ell'=\ell+1$ in
(\ref{dq1}), we first use Lemma 2(i) \cite{APsu3} or (III.36); next
we extend the interval of summation to $1\le\ell\le m_Q$ and
use (\ref{sq}); lastly we use the identities (III.55) and
(\ref{beta0}), to obtain
\begin{eqnarray}
d_{m,Q}&&=
(\Lambda^Q_{0})\vf^{-1}\sum_{\ell=1}^{m}S^Q_{m-\ell}\,\ell\Lambda^Q_\ell
\label{dq12}\\
&&=(\Lambda^Q_{0})\vf^{-1}\sum_{j=1}^{m_Q}\beta^Q_{j,0}z_{j,Q}^{1-m}
\sum_{\ell=1}^{m_Q} \ell \Lambda^Q_\ell z_{j,Q}^{\ell-1}
=-\sum_{j=1}^{m_Q} z_{j,Q}^{-m}.
\label{dq2}
\end{eqnarray}
This then generalizes (II.A.3). Using (\ref{dq}) followed by
(\ref{invxm}), (\ref{invxp}), (\ref{ocbk}), (\ref{cbok})
and (III.36) of Lemma 2 again, we find
\begin{eqnarray}
\sum_{n=1}^{m}\Lambda^Q_{m-n}d_{n,Q}
&&=\sum_{n=0}^{m-1}\Lambda^Q_{m-1-n}
\langle\Omega|{\bf x}^{+}_{n,Q}{\bf x}^{-}_{1,Q}|\Omega\rangle
\nonumber\\
&&=\frac{\omega^{2Q}}{\Lambda^Q_0}
\langle\Omega|{\bfC}_0^{(m N+Q)}{\bfB}_1^{(m N-N+Q)}
{\bfC}_0^{(Q)}{\bfB}_1^{(N+Q)}|\Omega\rangle
\nonumber\\
&&=(\Lambda^Q_0)\vf^{-1}
\sum_{{\{0\le n_j\le N-1\}}\atop{n_1+\cdots+n_L=N}}
\bK^{\vp}_{mN-N+Q}(\{n_j\}){K}^{\vp}_Q(\{n_j\})
\nonumber\\
&&=m\Lambda^Q_{m},
\label{dlambda}
\end{eqnarray}
generalizing (II.A.1). Next, we shall show
\begin{equation}
d_{m,Q}=\langle\Omega|{\bf h}_{m,Q}|\Omega\rangle
=\langle\Omega|{\bf x}^{+}_{m-k,Q}
{\bf x}^{-}_{k,Q}|\Omega\rangle,\quad\hbox{for }\;1<k\le m,
\label{dq3}
\end{equation}
which is a necessary condition that the loop algebra or subalgebra
exists. Again we substitute (\ref{xm}) and (\ref{xp}) into the
right hand side of the above equation, then use (\ref{ocbk}) and
(\ref{cbok}), and finally use (III.37), which is Lemma 2(ii)
in \cite{APsu3}, to find
\begin{eqnarray}
\langle\Omega|{\bf x}^{+}_{m-k,Q}{\bf x}^{-}_{k,Q}|\Omega\rangle
&&=(\Lambda^Q_{0})\vf^{-2}\sum_{\ell=0}^{m-k} S^Q_{m-k-\ell}
\sum_{n=0}^{k-1} S^Q_{k-1-n}
\nonumber\\
&&\qquad\qquad\times\sum_{j=0}^\ell
(n-\ell+1+2j)\Lambda^Q_{\ell-j}\Lambda^Q_{n+1+j}.
\label{dq4}
\end{eqnarray}
In the last step we have used the symmetry
$\Theta_{\ell,m,k}=\Theta_{m,\ell,k}$ for the quantity in (III.38)
studied in Lemma 2. This is a direct consequence of the identities
\begin{equation}
K_m(\{n_{L+1-j}\})=\bar K_m(\{n_j\}),\quad
N_{L+1-j}(\{n_{L+1-\ell}\})=\bar N_j(\{n_{\ell}\}),
\end{equation}
for the quantities defined in (III.7) and (III.8).

Interchanging the order of summation over $\ell$ with the one
over $j$ and then letting $\ell'=\ell-j$, we find that the summation
over $\ell'$ can be carried out by using (\ref{lsq}) and (\ref{dq12}),
after observing that the term proportional to $\ell'$ vanishes for
$j=m-k$.
We obtain
\begin{eqnarray}
\fl\langle\Omega|{\bf x}^{+}_{m-k,Q}{\bf x}^{-}_{k,Q}|\Omega\rangle
&&=(\Lambda^Q_{0})\vf^{-2}\sum_{j=0}^{m-k}
\sum_{n=0}^{k-1} S^Q_{k-1-n}\Lambda^Q_{n+1+j}
\nonumber\\
&&\qquad\qquad\times\sum_{\ell'=0}^{m-k-j}(n+1+j-\ell')
S^Q_{m-k-j-\ell'}\Lambda^Q_{\ell'}
\nonumber\\
&&=(\Lambda^Q_{0})\vf^{-1}\bigg[\sum_{j=0}^{m-k}
\sum_{n=0}^{k-1}S^Q_{k-1-n}\Lambda^Q_{n+1+j}(n+1+j)\delta\vp_{m,k+j}
\nonumber\\
&&\qquad\qquad-\sum_{j=0}^{m-k-1}\sum_{n=0}^{k-1}
S^Q_{k-1-n}\Lambda^Q_{n+1+j}d\vp_{m-k-j,Q}\bigg].
\label{dq5}
\end{eqnarray}
We then let $n\to k-1-n$ and $j\to m-k-j$, resulting in
\begin{eqnarray}
\fl\langle\Omega|{\bf x}^{+}_{m-k,Q}{\bf x}^{-}_{k,Q}|\Omega\rangle
&&= 
(\Lambda^Q_{0})\vf^{-1}\bigg[\sum_{n=0}^{k-1}S^Q_n\Lambda^Q_{m-n}(m-n)
-\sum_{j=1}^{m-k}\sum_{n=0}^{k-1}
S^Q_n\Lambda^Q_{m-j-n}d\vp_{j,Q}\bigg]
\nonumber\\
&&=(\Lambda^Q_{0})\vf^{-1}\bigg[\Lambda^Q_{0}d\vp_{m,Q}
-\sum_{n=k}^{m} S^Q_n\Lambda^Q_{m-n}(m-n)
\nonumber\\
&&\qquad\qquad-\sum_{j=1}^{m-k}d\vp_{j,Q}
\Big(\Lambda^Q_{0}\delta\vp_{m,j}
-\sum_{n=k}^{m-j} S^Q_n\Lambda^Q_{m-j-n}\Big)
\bigg]=d\vp_{m,Q},
\label{dq6}
\end{eqnarray}
where (\ref{dq12}) is used for the first sum and (\ref{lsq}) for the
second sum. Since $k>1$, we find $\delta_{m,j}=0$ for $1\le j\le m-k$.
Finally, after first interchanging the sums
$\sum_{j=1}^{m-k}\sum_{n=k}^{m-j}=\sum_{n=k}^{m-1}\sum_{j=1}^{m-n}$,
(\ref{dlambda}) is used to show that (\ref{dq3}) holds
for $1\le k\le m$, but not for $k<1$ when $Q\ne0$, as the
${\bf x}^{-}_{k,Q}|\Omega\rangle$ in (\ref{xm}) are defined only
for $k\ge 1$, while the $\langle\Omega|{\bf x}^{+}_{k,Q}$ in (\ref{xp})
are given for $k\ge0$. Thus, this shows that only a subalgebra may
exist, like those discussed in \cite{Degu3}.

%%%%%%%%%%%%%%%%%%%%%%%%%%%%%%%%%%%%%%%%%%%%%%%%%%%%%%%%%%%%%%%%%%%%%%%%
\subsection{Generators of the quantum loop subalgebra\label{loop}}

Formulae (\ref{xm}) for $n=1$ and (\ref{xp}) for $n=0$ suggest that%
\footnote{Equation (\ref{xmpo}) is similar to (3.43) in \cite{DFM}
for the XXZ model at roots of unity.}
\begin{equation}
{\bf x}^{-}_{1,Q}=
(\omega^{Q}/\Lambda^Q_{0}){\bfC}_0^{(Q)}{\bfB}_1^{( N+Q)},\quad
{\bf x}^{+}_{0,Q}=
(\omega^{Q}/\Lambda^Q_{0}){\bfC}_0^{(N+Q)}{\bfB}_1^{(Q)},
\label{xmpo}
\end{equation}
and
\begin{equation}
\fl{\bf h}\vp_{1,Q}=[{\bf x}^{+}_{0,Q},{\bf x}^{-}_{1,Q}],\quad
{\bf x}^{-}_{n+2,Q}=\half[{\bf h}\vp_{1,Q},{\bf x}^{-}_{n+1,Q}],\quad
{\bf x}^{+}_{n+1,Q}=-\half[{\bf h}\vp_{1,Q},{\bf x}^{+}_{n,Q}],
\label{xmph}
\end{equation}
for $0\le n\le \infty$. Because of the complex form of these
operators, to prove the Serre relations
\begin{equation}
[[[{\bf x}^{+}_{0,Q},{\bf x}^{-}_{1,Q}],
{\bf x}^{-}_{1,Q}],{\bf x}^{-}_{1,Q}]=0,
\qquad
[{\bf x}^{+}_{0,Q},[{\bf x}^{+}_{0,Q},
[{\bf x}^{+}_{0,Q},{\bf x}^{-}_{1,Q}]]]=0
\label{serre}
\end{equation}
is highly nontrivial. We can prove by induction the following,
\begin{eqnarray}
({\bf x}^{-}_{1,Q})^n|\Omega\rangle=
n!(\omega^{Q}/\Lambda^Q_{0}){\bfC}_0^{(Q)}{\bfB}_1^{( nN+Q)}
|\Omega\rangle,\quad
1\le n\le m_Q,\label{nxm}\\
\fl({\bf x}^{+}_{0,Q})^m({\bf x}^{-}_{1,Q})^n|\Omega\rangle
=m!n!(\omega^{Q}/\Lambda^Q_{0}){\bfC}_0^{(mN+Q)}
{\bfB}_1^{( nN+Q)}|\Omega\rangle,\quad
0\le m\le n\le m_Q.
\label{mnxpxm}
\end{eqnarray}
The proofs are left to Appendix A.

These relations can be used to show that the first Serre
relation in (\ref{serre}) holds for
$({\bf x}^-_{1,Q})^n|\Omega\rangle$. That is
\begin{equation}
[[[{\bf x}^{+}_{0,Q},{\bf x}^{-}_{1,Q}],{\bf x}^{-}_{1,Q}]
{\bf x}^{-}_{1,Q}]({\bf x}^-_{1,Q})^{n}|\Omega\rangle
=0.
\label{pfserre}
\end{equation}
These details are in Appendix B.
We managed to show that it also holds for
$({\bf x}^+_{0,Q})({\bf x}^-_{1,Q})^n|\Omega\rangle$, but we have
been unable to prove it for
$({\bf x}^+_{0,Q})^m({\bf x}^-_{1,Q})^n|\Omega\rangle$ for $m>1$.
Moreover, even if one would prove (\ref{serre}) on these states,
this would still by far not enough.

For general states $|\{n_j\}\rangle$ satisfying the cyclic boundary
condition $n_1+\cdots+n_L\equiv0\;(\hbox{mod}\,N)$, we again tested
the Serre relation for small systems on a computer using Maple.
The simplest nontrivial cases are $N=3$, $L=6$ and $n_1+\cdots+n_6=3$.
Yet compared with the case $Q=0$, the complexity increases enormously;
each case, running in Maple 12 on ANU computers in Theoretical Physics
took five days. We have found that the Serre relation holds for
all cases tested. Even though a formal proof is still lacking,
we believe that the Serre relation (\ref{serre}) holds.
As a consequence, we believe that the following loop subalgebra holds,
\begin{eqnarray}
{\bf h}_{n,Q}=[{\bf x}^{+}_{n-k,Q},{\bf x}^{-}_{k,Q}],\quad
\hbox{for}\quad 1\le k\le n,
\nonumber\\
\fl{\bf x}^{-}_{n+k+1,Q}=\half[{\bf h}_{n,Q},{\bf x}^{-}_{k+1,Q}],
\quad
{\bf x}^{+}_{n+k,Q}=-\half[{\bf h}_{n,Q},{\bf x}^{+}_{k,Q}],
\quad n>1,\; k>0.
\label{xmphg}
\end{eqnarray}
Since the indices here are nonnegative integers only, this is not the
entire loop algebra, but a subalgebra as in \cite{Degu3}.

%%%%%%%%%%%%%%%%%%%%%%%%%%%%%%%%%%%%%%%%%%%%%%%%%%%%%%%%%%%%%%%%%%%%%%%%
\subsection{Generators of the ${\mathfrak{sl}}_2$ algebra}

In (\ref{Epm}), the generators ${\bf E}^{\pm}_{m,Q}$ on
the ground state were given, but this is not sufficient.
We can now define them in terms of generators
of the loop algebra as in (II.13), namely
\begin{equation}
{\bf E}_{m,Q}^{+}=
\sum_{n=0}^{m_Q-1}\beta^{\ast Q}_{m,n}z\vp_{m,Q}{\bf x}_{n+1,Q}^{-},
\quad
{\bf E}_{m,Q}^{-}=-
\sum_{n=0}^{m_Q-1}\beta^{\ast Q}_{m,n}{\bf x}_{n,Q}^{+}.
\label{epem}
\end{equation}
Here $\beta^{\ast Q}_{m,n}$ is defined through (\ref{beta}) with
$z_{k,Q}$ replaced by $1/z_{k,Q}$, i.e.
\begin{equation}
f^{\ast Q}_j(z)=\prod_{\ell\ne j}{\frac{z-z^{-1}_{\ell,Q}}
{z^{-1}_{j,Q}-z^{-1}_{\ell,Q}}}=
\sum_{n=0}^{m_Q-1}\beta^{\ast Q}_{j,n}z^n, \quad
f^{\ast Q}_j(z^{-1}_{k,Q})=\delta\vp_{j,k}\,,
\label{betastar}
\end{equation}
so that (\ref{vdm}) is now replaced by
\begin{equation}
\sum_{n=0}^{m_Q-1}\beta^{\ast Q}_{j,n}z^{-n}_{k,Q}=\delta_{j,k},\qquad
\sum_{k=1}^{m_Q}z^{-n}_{k,Q}\beta^{\ast Q}_{k,m}=\delta_{n,m},
\quad(0\le n\le m_Q-1).
\label{vdmstar}
\end{equation}
The difference in the two equations in (\ref{epem}) is due to the
fact that ${\bf x}_{n,Q}^{-}$ is defined for $n\ge1$, while
${\bf x}_{n,Q}^{+}$ is defined for $n\ge0$. We can also define
\begin{equation}
{\bf H}\vp_{m,Q}=
\sum_{n=0}^{m_Q-1}\beta^{\ast Q}_{m,n}z\vp_{m,Q}{\bf h}\vp_{n+1,Q}.
\label{eH}
\end{equation}
Using (\ref{vdmstar}) we can invert (\ref{epem}) and (\ref{eH}) as
\begin{equation}
\fl{\bf x}^{-}_{n,Q}=\sum_{m=1}^{m_Q}z^{-n}_{m,Q}{\bf E}_{m,Q}^{+},\quad
{\bf x}^{+}_{n,Q}=-\sum_{m=1}^{m_Q} z^{-n}_{m,Q}{\bf E}_{m,Q}^{-},\quad
{\bf h}\vp_{n,Q}=\sum_{m=1}^{m_Q} z^{-n}_{m,Q}{\bf H}\vp_{m,Q}.
\label{epex}
\end{equation}
consistent with (\ref{xm}) and (\ref{xp}) for the action on the
ground state $|\Omega\rangle$. Replacing $n$ by $n+\ell$ in
(\ref{epex}) and then inverting back using (\ref{vdmstar}) we find
\begin{eqnarray}
&&{\bf E}_{m,Q}^{+}=
\sum_{n=0}^{m_Q-1}\beta^{\ast Q}_{m,n}
z^{\ell}_{m,Q}{\bf x}_{n+\ell,Q}^{-},
\quad
{\bf E}_{m,Q}^{-}=-
\sum_{n=0}^{m_Q-1}\beta^{\ast Q}_{m,n}
z^{\ell}_{m,Q}{\bf x}_{n+\ell,Q}^{+},\nonumber\\
&&{\bf H}\vp_{m,Q}=
\sum_{n=0}^{m_Q-1}\beta^{\ast Q}_{m,n}
z^{\ell}_{m,Q}{\bf h}\vp_{n+\ell,Q}.
\label{epemh}
\end{eqnarray}
generalizing (II.13), but only for $\ell\ge1$ or $0$.
From (\ref{epemh}) we can derive the usual $\otimes\mathfrak{sl}_2$
commutation relations as in (II.15), for example
\begin{eqnarray}
&&[{\bf E}_{m,Q}^{+},{\bf E}_{j,Q}^{-}]=
\sum_{n=0}^{m_Q-1}\beta^{\ast Q}_{m,n}z\vp_{m,Q}
\sum_{k=0}^{m_Q-1}\beta^{\ast Q}_{j,k}\,
[{\bf x}_{n+1,Q}^{-},{\bf x}_{k,Q}^{+}]\nonumber\\
&&\quad=\sum_{n=0}^{m_Q-1}\beta^{\ast Q}_{m,n}z\vp_{m,Q}z^{-n-1}_{j,Q}
\sum_{k=0}^{m_Q-1}\beta^{\ast Q}_{j,k}
z^{n+1}_{j,Q}{\bf h}\vp_{k+n+1,Q}
=\delta\vp_{m,j}{\bf H}\vp_{j,Q}
\label{H}
\end{eqnarray}
follows after using (\ref{epem}), (\ref{xmphg}), (\ref{epemh})
and (\ref{vdmstar}) in order.

Because of equation (\ref{mnxpxm}), we may rewrite (\ref{epo}) as
\begin{equation}
\fl{\bf E}_{m,Q}^+|\Omega\rangle
=\beta^Q_{m,0}\sum_{\ell=1}^{m_Q} z_{m,Q}^{\ell}
({\bf x}_{0,Q}^+)^{(\ell-1)}
({\bf x}_{1,Q}^-)^{(\ell)}|\Omega\rangle,\quad 
({\bf x}_{n,Q}^{\pm})^{(\ell)}
\equiv\frac{({\bf x}_{n,Q}^{\pm})^\ell}{\ell!}.
\label{epo1}
\end{equation}
Assuming that the Serre relation (\ref{serre}) holds, we may
again prove by induction
\begin{eqnarray}
&&[({\bf x}_{0,Q}^+)^{(j)},({\bf x}_{k,Q}^-)]
=({\bf x}_{0,Q}^+)^{(j-1)}{\bf h}\vp_{k,Q}-
{\bf x}_{k,Q}^+({\bf x}_{0,Q}^+)^{(j-2)},
\nonumber\\
&&[({\bf x}_{k,Q}^+),({\bf x}_{1,Q}^-)^{(j)}]
=({\bf x}_{1,Q}^-)^{(j-1)}{\bf h}\vp_{k+1,Q}+
{\bf x}_{k+2,Q}^-({\bf x}_{1,Q}^-)^{(j-2)},
\nonumber\\
&&\fl[{\bf h}\vp_{k,Q},({\bf x}_{0,Q}^+)^{(j)}]
=-2{\bf x}_{k,Q}^+({\bf x}_{0,Q}^+)^{(j-1)},
\quad [{\bf h}\vp_{k,Q},({\bf x}_{1,Q}^-)^{(j)}]=
2{\bf x}_{k+1,Q}^-({\bf x}_{1,Q}^-)^{(j-1)},
\label{commuta}
\end{eqnarray}
so that Appendix B in \cite{APsu2} can be repeated here to show that
\begin{eqnarray}
\fl{\bf E}_{j,Q}^{+}{\bf E}_{m,Q}^{+}|\Omega\rangle
=&&\beta^Q_{m,0}\Bigl\{
(1-z\vp_{m,Q}/z\vp_{j,Q})^2\sum_{\ell=1}^{m_Q-1}
z_{m,Q}^{\ell}({\bf x}_{0,Q}^+)^{(\ell-1)}
({\bf x}_{1,Q}^-)^{(\ell)}{\bf E}_{j,Q}^{+}|\Omega\rangle
\nonumber\\
&&+(1-z\vp_{m,Q}/z\vp_{j,Q})\sum_{\ell=2}^{m_Q}
z_{m,Q}^{\ell}({\bf x}_{0,Q}^+)^{(\ell-2)}
({\bf x}_{1,Q}^-)^{(\ell)}|\Omega\rangle\Bigr\}.
\label{eeo2}
\end{eqnarray}
Again, we have $({\bf E}_{m,Q}^{+})^2|\Omega\rangle=0$.

If we let 
\begin{equation}
{\bf x}^{-}_{0,Q}=(\Lambda^Q_{0})\vf^{-1}
{\bfC}_{L-1}^{(Q)}{\bfB}_L^{( N+Q)},\quad
{\bf x}^{+}_{-1,Q}
=(\Lambda^Q_{0})\vf^{-1}{\bfC}_{L-1}^{(N+Q)}{\bfB}_L^{(Q)},
\label{xmxpng}
\end{equation}
so that for $Q=0$ we have ${\bf x}^{-}_{0,Q}\to{\bf x}^{-}_{0}$
and ${\bf x}^{+}_{-1,Q}\to{\bf x}^{+}_{-1}$, one can see from 
(\ref{tauBC1L}) that ${\bf x}^{-}_{0,Q}|\Omega\rangle$
and ${\bf x}^{+}_{-1,Q}|\Omega\rangle$ are not eigenvectors
of $\tau_2(t_q)|_Q$, but eigenvectors of $\tau_2(t_q)|_{N-Q}$.
However,
\begin{equation}
\fl{\bf x}^{-}_{0,N-Q}
=(\Lambda^{N-Q}_{0})\vf^{-1}{\bfC}_{L-1}^{(N-Q)}
{\bfB}_L^{( 2N-Q)},\quad
{\bf x}^{+}_{-1,N-Q}
=(\Lambda^{N-Q}_{0})\vf^{-1}{\bfC}_{L-1}^{(2N-Q)}{\bfB}_L^{(N-Q)},
\label{xmxpng2}
\end{equation}
and their products when applied to $|\Omega\rangle$ give eigenvectors
of $\tau_2(t_q)|_Q$, but corresponding to the Drinfeld polynomial 
$P_{N-Q}(z)$. It is possible to express the ${\bf E}_{j,Q}^{\pm}$
also in terms of these operators.

%%%%%%%%%%%%%%%%%%%%%%%%%%%%%%%%%%%%%%%%%%%%%%%%%%%%%%%%%%%%%%%%%%%%%%%%
\section{Transfer Matrix Eigenvectors}

From (\ref{eigvec1}) we see that $2^{m_Q}$ eigenvectors of $\tau_2$
obeying (\ref{ground1}) can be generated by operating the $m_Q$
operators ${\bf E}_{j,Q}^{+}$ on the ground state $|\Omega\rangle$,
while $2^{m_Q}$ eigenvectors satisfying (\ref{ground2}) are found by
operating the $m_Q$ operators ${\bf\bE}_{j,N-Q}^{-}$ on
$|\bar\Omega\rangle$. The ${\bf\bE}_{j,N-Q}^{-}$ differ from the
${\bf E}_{j,N-Q}^{-}$ in that the positions of the ${\bfB}_1$
and ${\bfC}_0$ are interchanged, as can be seen from (\ref{eigvec1})
and (\ref{eigvec2}). We now show how the transfer matrices of the
superintegrable chiral Potts model in the corresponding sectors can be
expressed in terms of these generators and how the resulting $2^r$
eigenvectors can be obtained. (Here $r=m_Q+1$ for $Q\ne0$.)

%%%%%%%%%%%%%%%%%%%%%%%%%%%%%%%%%%%%%%%%%%%%%%%%%%%%%%%%%%%%%%%%%%%%%%%%
\subsection{Ground state sector eigenvalues for $Q\ne0$}

From (6.2) and (6.14) of Baxter \cite{BaxIf1}, we find%
\footnote{We have chosen the multiplication of transfer matrices up
to down, rather than down to up, making our transfer matrices the
transposes of those of Baxter. Therefore, in (I.15) the operator
$\mathbf{X}$ is the inverse of the one used by Baxter, so that
comparing with \cite{BaxIf1} we need to replace $Q\to N-Q$, when
$Q\ne0$.}
\begin{equation}
\mathcal{T}_Q(x_q,y_q){\bf x}=x_q^{P_a}y_q^{P_b}
\mathcal{G}(\lambda_q){\bf y},
\label{gg1}
\end{equation}
where $P_a=Q$ and $P_b=0$ for the $2^{m_Q}$ eigenvectors satisfying
(\ref{eigv1}), while $P_a=0$ and $P_b=N-Q$ for the $2^{m_Q}$
eigenvectors obeying (\ref{eigv2}). Thus, comparing with (II.4)
for $Q=0$ and (6.24) and (6.25) of \cite{BaxIf1} with $F\equiv1$
for general $Q$, we have
\begin{equation}
\mathcal{G}\vp_a(\lambda\vp_q)\mathcal{G}\vp_a(\lambda_q^{-1}) 
=Nt_p^{rN}P\vp_{Q}(t^N)
=Nt_p^{rN}\Lambda^{Q}_{ m_Q}
\prod_{j=1}^{m_Q}[(t\vp_q/t\vp_p)^N-z\vp_{j,Q}]
\label{gg2}
\end{equation}
for the former case, and 
\begin{equation}
\fl{\mathcal{G}\vp_b}(\lambda\vp_q){\mathcal{G}\vp_b}(\lambda_q^{-1}) 
=\omega^{Q}Nt_p^{rN}P\vp_{N-Q}(t^N)
=\omega^{Q}Nt_p^{rN}\Lambda^{Q}_0
\prod_{j=1}^{m_Q}[(t\vp_q/t\vp_p)^N-z^{-1}_{j,Q}]
\label{gg3}
\end{equation}
for the latter. Here subscripts $a$ and $b$ have been inserted to
distinguish the two cases and $rN=(N-1)L$.
Because $\Lambda^{Q}_{m_Q-j}=\Lambda^{N-Q}_{j}$, the roots of
$P_{N-Q}(z)$ are the inverses of the roots of $P_{Q}(z)$.
Consequently, as in (II.8), we may write
\begin{equation}
\fl\mathcal{G}_a(\lambda_q)
={D_Q}\prod_{j=1}^{m_Q}(A_{j,Q}\pm B_{j,Q}),\quad
{\mathcal{G}_b}(\lambda_q)
={\hat D}_Q\prod_{j=1}^{m_Q}(A_{j,N-Q}\pm B_{j,N-Q}),
\label{eigent}
\end{equation}
where
\begin{eqnarray}
A\vp_{j,Q}=\cosh\theta\vp_{j,Q}(1-\lambda_q^{-1}),\quad
B\vp_{j,Q}= \sinh \theta\vp_{j,Q}(1+\lambda_q^{-1}),
\label{abj}\\
D\vp_{Q}=
(Nt_p^{N}\Lambda^{N-Q}_{0})^{\frac12}({k'}/{k^2})^{{\frac12}m_Q},
\quad
{\hat D}\vp_Q=
(\omega^{Q}Nt_p^{N}\Lambda^{Q}_0)^{\frac12}({k'}/{k^2})^{{\frac12}m_Q},
\label{Dq}
\end{eqnarray}
with $\theta_{j,Q}$ given by (II.6) replacing
$z_j\to z_{j,Q}$, i.e.
\begin{equation}
\fl2 \cosh 2\theta\vp_{j,Q}
=k'+k'^{-1}-k^2 t_p^Nz\vp_{j,Q}/k',\quad
\theta\vp_{j,N-Q}=\theta^\ast_{j,Q},
\quad z\vp_{j,Q}z^\ast_{j,Q}=1.
\label{theta}
\end{equation}
We have also changed $A_{j,Q}$ compared with \cite{APsu2} by dropping
the constant $\rho$, absorbing it into the constant $D_Q$ instead.

%%%%%%%%%%%%%%%%%%%%%%%%%%%%%%%%%%%%%%%%%%%%%%%%%%%%%%%%%%%%%%%%%%%%%%%%
\subsection{Erratum and added details for \cite{APsu2}.\label{errors}}

In \cite{APsu2}, we have shown---comparing (II.40) and (II.42)---that 
\begin{equation}
\langle{\bar\Omega}|\mathcal{T}_Q(x_q,y_q)|\Omega\rangle
=\langle\Omega|\mathcal{T}_Q(x_q,y_q)|{\bar\Omega}\rangle.
\label{otbo}
\end{equation}
Also, from (II.81) we have
\begin{equation}
\langle{\bar\Omega}|=\langle{\Omega}|\prod_{j=1}^r{\bf E}_j^{-},\quad
|{\bar\Omega}\rangle
=\prod_{j=1}^r{\bf E}_j^{+}|{\Omega}\rangle,
\label{obo}
\end{equation}
with $r=m_0$. To satisfy (\ref{otbo}), we have made in \cite{APsu2}
the assumption (II.43) that the transfer matrix is of the form
\begin{equation}
{\cal T}_0(x_q,y_q)=\prod_{j=1}^r[X\vp_j-Y\vp_j{\bf H}\vp_j+
({\bf E}^+_j +{\bf E}^-_j)Z\vp_j].
\label{transfer1e}
\end{equation}
However, the condition in (\ref{otbo}) can still hold if the
transfer matrix takes the form
\begin{equation}
{\cal T}_0(x_q,y_q)=\prod_{j=1}^r[X\vp_j-Y\vp_j{\bf H}\vp_j+
Z\vp_j{\bf E}^+_j +{\hat Z}\vp_j{\bf E}^-_j],
\label{transfer1f}
\end{equation}
as long as $\prod_{j=1}^r Z_j= \prod_{j=1}^r{\hat Z}_j$.
Instead of (II.83), this more general form yields
\begin{equation}
\frac{X_m-Y_m}{{\hat Z}_m}=
\frac{x_p^N-y_q^N z_m^{-1}}{x_p^N-y_q^N},
\label{ratio2}
\end{equation}
as we must then replace $Z_m$ by ${\hat Z}_m$ in (II.82).
The transfer matrices have the symmetry
${\cal T}_0(x_q,y_q)\leftrightarrow{\hat\mathcal{T}}_0(x_q,y_q)$
under the interchange $p\leftrightarrow p'$.
For ${\hat\mathcal{T}}_0(x_q,y_q)$, the ratios are given in (II.99).
Substituting $x_p\to y_p$ in (\ref{ratio2}) because of $p\to p'$,
and comparing the resulting equation with (II.99), we find the
necessary condition ${\hat Z}_j=-z_j Z_j$. Since
$\prod_{j=1}^r(-z_j)=1$, such a choice for the transfer matrix still
satisfies (\ref{otbo}). For this choice, the determinantal condition
(II.86) becomes $X_j^2-Y_j^2-Z\vp_j{\hat Z}\vp_j=A_j^2-B_j^2$, with
the result for $X_j-Y_j$ in (II.85) multiplied by $-z_j$, and it
can be solved using $\lambda\vp_p\equiv\mu_p^N$, (I.2), (I.5) and
(II.87) as
\begin{equation}
{\bar\epsilon}_j^2\equiv\epsilon_j^2/\rho^2=1/[k'(z_j-1)\lambda_p],
\label{bareps}
\end{equation}
which differs from (II.89) by a factor $(-z_j)^{-1}$.
Next, (II.91) is altered to become
\begin{eqnarray}
\fl m_{11}={\bar\epsilon}_j k'\lambda_p,\quad
m_{21}={\bar\epsilon}_j k'\lambda_p z_j,\quad
m_{12}=-{\bar\epsilon}_j k'\lambda_p ,\quad
m_{22}={\bar\epsilon}_j(z_j-1-k'z_j\lambda_p),\nonumber\\
\fl n_{11}=-{\bar\epsilon}_j(\lambda_p-z_j\lambda_p+k'z_j),\quad
n_{21}=-{\bar\epsilon}_j k'z_j,\quad
n_{12}=n_{22}={\bar\epsilon}_j k',
\label{mnij1}
\end{eqnarray}
so that $m_{21}=-z_j m_{12}$ and $n_{21}=-z_j n_{12}$. Comparing with
(II.91), we find
\begin{eqnarray}
&&m_{l,1} \to (-z_j)^{\halfs}m_{l,1},\;\;\quad
n_{l,1}\to (-z_j)^{\halfs}n_{l,1},\nonumber\\
&&m_{l,2}\to (-z_j)^{-\halfs}m_{l,2},\quad
n_{l,2}\to (-z_j)^{-\halfs} n_{l,2},\quad\mbox{for }l=1,2.
\label{new}
\end{eqnarray}
Most of the other equations in \cite{APsu2} still hold, except for
a few modifications. One can easily show that the first two
equations in (II.C.7) become
\begin{equation}
\frac{r\vp_{21}}{r\vp_{11}}=-\frac{T^{\ast}_{21}}{T^{\ast}_{22}}=
z\vp_j\frac{s\vp_{12}}{s\vp_{22}},\qquad
\frac{r\vp_{12}}{r\vp_{22}}=-\frac{T\vp_{22}}{T\vp_{21}}=
\frac{s\vp_{21}}{z\vp_j s\vp_{11}},
\label{Rj2}
\end{equation}
where
\begin{equation}
T\vp_{lk}=m\vp_{lk}\e^{-\theta_j}+n\vp_{lk}\e^{\theta_j},\qquad
T^{\ast}_{lk}=m\vp_{lk}\e^{\theta_j}+n\vp_{lk}\e^{-\theta_j},
\label{tlk}
\end{equation}
as in (II.C.10) without the symmetry conditions, as now
$T\vp_{21}=-z\vp_jT\vp_{12}$, $T^{\ast}_{21}=-z\vp_jT^{\ast}_{12}$
from (\ref{new}). This same (\ref{new}) leaves equations (II.C.11)
unchanged.

We can now solve (II.C.6) and (\ref{Rj2}) as
\begin{eqnarray}
&\mathcal{S}_j=\left(\begin{array}{cc}
\displaystyle
\frac{\e^{2\theta_j}-k'}{2\sinh(2\theta_j)s\vp_{22_{\strut}}}
&\displaystyle\frac{(\lambda_p-k')s\vp_{22}}{\e^{2\theta_j}-k'}\\
\displaystyle\frac{(\e^{2\theta_j}-k')(\e^{-2\theta_j}-k')}
{2\sinh(2\theta_j)(\lambda_p-k')s\vp_{22}}
&s\vp_{22}\\
\end{array}\right),
\label{sjuneq}
\\
\nonumber\\
&\mathcal{R}_j=\left(\begin{array}{cc}
\displaystyle
\frac{\e^{2\theta_j}-k'}{2\sinh(2\theta_j)r\vp_{22_{\strut}}}
&\displaystyle\frac{(\lambda^{-1}_p-k')r\vp_{22}}{\e^{2\theta_j}-k'}\\
\displaystyle\frac{(\e^{2\theta_j}-k')(\e^{-2\theta_j}-k')}
{2\sinh(2\theta_j)(\lambda^{-1}_p-k')r\vp_{22}}
&r\vp_{22}\\
\end{array}\right),
\label{rjuneq}
\end{eqnarray}
where we have used (\ref{mnij1}) and (\ref{tlk})
and we have eliminated $z_j$ using
\begin{equation}
z_j=\frac{(\e^{2\theta_j}-k')(\e^{-2\theta_j}-k')}
{(1-k'\lambda\vp_p)(1-k'\lambda_p^{-1})},
\label{ljtjtozj}
\end{equation}
following from (II.5) and (II.6). From (II.C.9), (\ref{mnij1}) and
(\ref{bareps}) we find one relation between $r\vp_{22}$
and $s\vp_{22}$,
\begin{equation}
r\vp_{22}=\pm s\vp_{22}\,
\sqrt{\frac{-(\e^{2\theta_j}-\lambda_p)(\lambda_p-k')}
{(\e^{2\theta_j}-\lambda^{-1}_p)(1-k'\lambda_p)}}.
\label{rjtosj}
\end{equation}
Note that interchanging $\lambda\vp_p$ and
$\lambda\vp_{p'}=\lambda^{-1}_p$
interchanges $r\vp_{22}$ and $s\vp_{22}$ in (\ref{rjtosj}) and
$\mathcal{R}_j$ and $\mathcal{S}_j$ in (\ref{sjuneq}) and
(\ref{rjuneq}), which is a required symmetry. The original choice
in \cite{APsu2} does not satisfy this property and is incorrect.
Equation (II.94), fixing the remaining free parameter $s\vp_{22}$ and
the remaining $\pm$ sign in (\ref{rjtosj}) by a special choice,
needs be changed to
\begin{equation}
r\vp_{22}=s\vp_{11},\quad r\vp_{21}=z\vp_j s\vp_{12},\quad
r\vp_{12}=z^{-1}_j s\vp_{21},\quad r\vp_{11}=s\vp_{22}.
\label{Rij}
\end{equation}

For the special case $p=p'$ we must have $\lambda_p=\pm1$. If
$\lambda_p=+1$, we find from (\ref{rjtosj}) that
$r\vp_{22}=\pm\mathrm{i}s\vp_{22}$, so that
$\mathcal{R}_j=\pm\mathrm{i}\,\mathcal{S}_j\sigma^z=
\pm\mathrm{i}\,\mathcal{S}_j{\mbox{\tiny$\pmatrix{1&0\cr0&-1}$}}$.
On the other hand, if $\lambda_p=-1$, we find
$r\vp_{22}=\pm s\vp_{22}$ and $\mathcal{R}_j=\pm\mathcal{S}_j$.

%%%%%%%%%%%%%%%%%%%%%%%%%%%%%%%%%%%%%%%%%%%%%%%%%%%%%%%%%%%%%%%%%%%%%%%%
\subsection{Eigenvectors corresponding to
$x_q^Q\mathcal{G}^{}_a(\lambda^{}_q)$ \label{eigvec}}

We consider first the eigenvectors of the transfer matrix related to
(\ref{eigv1}) and (\ref{gg2}). On the corresponding vector subspace,
similar to (\ref{transfer1f}) and (II.26), we let
\begin{eqnarray}
{\cal T}\vp_Q(x\vp_q,y\vp_q)
&&=x_q^Q D\vp_Q\prod_{j=1}^{m_Q}
\left[X\vp_{j,Q}-{\bf H}\vp_{j,Q}Y\vp_{j,Q}+
({\bf E}^+_{j,Q} -z\vp_{j,Q}{\bf E}^-_{j,Q})Z\vp_{j,Q}\right]
\label{transfer1a}\\
&&=x_q^Q D\vp_Q\prod_{j=1}^{m_Q}{\bf\cal S}\vp_{j,Q}
(A\vp_{j,Q}-{\bf H}\vp_{j,Q} B\vp_{j,Q}){\bf\cal R}_{j,Q}^{-1},
\label{transfer1}
\end{eqnarray}
where ${\hat Z}_j=-z_j Z_j$ is inserted and
$\det{\bf\cal S}\vp_{j,Q}=\det{\bf\cal R}\vp_{j,Q}=1$. The $2^{m_Q}$
sought eigenvectors of the transfer matrix are given by
\begin{equation}
|{\bf \cal X}^Q_s\rangle=\prod_{j=1}^{m_Q}{\bf \cal R}\vp_{j,Q}
\prod_{m\in W_n}{\bf E}_{m,Q}^{+}|\Omega\rangle,\quad
|{\bf \cal Y}^Q_s\rangle=\prod_{j=1}^{m_Q}{\bf \cal S}\vp_{j,Q}
\prod_{m\in W_n}{\bf E}_{m,Q}^{+}|\Omega\rangle,
\label{evector}
\end{equation}
generalizing (II.28) to $Q\ne0$.
Here $s=\{s_1,s_2,\ldots,s_{m_Q}\}$, with $s_i=1$ if $i\in W_n$ and
$s_i=0$ if $i\not\in W_n$,
and $W_n=\{j_1,\ldots,j_n\}$, for $0\le n\le m_Q$, is any subset
of $\{1,2,\ldots,m_Q\}$, such that
\begin{equation}
{\cal T}\vp_Q(x\vp_q,y\vp_q)|{\bf \cal X}^Q_s\rangle=x_q^Q D\vp_Q
\prod_{j=1}^{m_Q}\left[ A\vp_{j,Q}+(-1)^{s_j}
B\vp_{j,Q}\right]|{\bf \cal Y}^Q_s\rangle.
\label{eig}
\end{equation}

To evaluate ${\bf \cal R}_{j,Q}$ and ${\bf \cal S}_{j,Q}$, we
start with (II.39), i.e.
\begin{equation}
\langle\Omega|\mathcal{T}\vp_Q(x\vp_q,y\vp_q)|\Omega\rangle
=N^{1-{\frac12}L}y_p^{rN}(x\vp_q/y\vp_p)^Q P\vp_Q(x^N_q/y^N_p).
\label{oto}
\end{equation}
We next use (II.37), (II.63) and $\sum_jN_j=\sum_j(L-j)n_j$ to obtain
\begin{equation}
\fl\langle \{n\vp_j\}|\mathcal{T}\vp_Q(x\vp_q,y\vp_q)|\Omega\rangle
=N^{1-{\frac 12}L}\omega^{-\sum_{j}j n_j}y_p^{rN}(1-x_q^N/y_p^N)
(x\vp_q/y\vp_p)^Q\,G\vp_Q(\{n\vp_j\},x_q^N/y_p^N),
\label{nto}
\end{equation}
from which, applying (\ref{ome2}) and (III.45), we find 
\begin{equation}
\fl\langle\Omega|{\bf E}_{m,Q}^{-}
\mathcal{T}\vp_Q(x\vp_q,y\vp_q)|\Omega\rangle
=-(\beta^Q_{m,0}/\Lambda_0^Q)N^{1-\frac 12L}y_p^{rN}(1-x_q^N/y_p^N)
(x\vp_q/y\vp_p)^Q\,{\mbox{\myeu h}}^Q_{m}(x^N_q/y^N_p).
\label{oepto3}
\end{equation}
Consequently, (III.57), (\ref{roots}) and (\ref{beta0}) can be used to
get the ratio
\begin{equation}
\frac{\langle \Omega|\mathcal{T}\vp_Q(x\vp_q,y\vp_q)|\Omega\rangle}
{\langle \Omega|{\bf E}_{m,Q}^{-}
\mathcal{T}\vp_Q(x\vp_q,y\vp_q)|\Omega\rangle}
=\frac{x_q^N-y_p^N z\vp_{m,Q}}{x_q^N-y_p^N}
=\frac{X\vp_{m,Q}+Y\vp_{m,Q}}{Z\vp_{m,Q}}.
\label{ratio1}
\end{equation}
Again, as in \cite{APsu2}, the ratio depends on $z_{m,Q}$ only, so
that ${\bf \cal R}_{m,Q}$ and ${\bf \cal S}_{m,Q}$ are independent
of the other roots of $P_Q(z)$. Since $|\bar\Omega\rangle$ and
$|\Omega\rangle$ are in different degenerate eigenspaces
of $\tau_2^Q$, we cannot evaluate the other ratios as in \cite{APsu2}.
However, we can consider now the alternate-row transfer matrix
${\hat\mathcal{T}}\vp_Q(y\vp_q,x\vp_q)$ at the special $q$-rapidity
with $x_q$ and $y_q$ interchanged. Of course, the fixed values $x_p$
and $y_p$ are also interchanged, as ${\hat\mathcal{T}}_Q$ has $p$ and
$p'$ reversed by definition. We then can write
\begin{eqnarray}
{\hat\mathcal{T}}\vp_Q(y\vp_q,x\vp_q)
&&=y_q^Q D\vp_Q\prod_{j=1}^{m_Q}
\left[{\bar X}\vp_{j,Q}-{\bf H}\vp_{j,Q}{\bar Y}\vp_{j,Q}+
({\bf E}^+_{j,Q}-z\vp_{j,Q}{\bf E}^-_{j,Q}){\bar Z}\vp_{j,Q}\right]
\nonumber\\
&&=y_q^Q D\vp_Q\prod_{j=1}^{m_Q}{\bf\cal R}\vp_{j,Q}
({\bar A}\vp_{j,Q}-{\bf H}\vp_{j,Q}{\bar B}\vp_{j,Q})
{\bf\cal S}_{j,Q}^{-1},
\label{transfer2}
\end{eqnarray}
with ${\bar A}_{j,Q}$ and ${\bar B}_{j,Q}$ obtained from $A_{j,Q}$ and
$B_{j,Q}$ in (\ref{abj}) replacing $\lambda_q^{-1}$ by $\lambda\vp_q$,
and we may use (II.95) with $n'_i\equiv0$ followed by (\ref{roots})
for $n_i\equiv0$, or by
(II.64) replacing
$N_i=\bar N_0-\bar N_{i-1}$, to find
\begin{equation}
\langle\Omega|\hat{\mathcal{T}}\vp_Q(y\vp_q,x\vp_q)|\Omega\rangle
=N^{1-{\frac12}L}x_p^{rN}(y\vp_q/x\vp_p)^Q P\vp_Q(y^N_q/x^N_p)
\label{ohto}
\end{equation}
and
\begin{equation}
\fl\langle\Omega|\hat{\mathcal{T}}\vp_Q
(y\vp_q,x\vp_q)|\{n\vp_j\}\rangle
=N^{1-{\frac 12}L}\omega^{\sum_{j}j n_j}x_p^{rN}(1-y_q^N/x_p^N)
(y\vp_q/x\vp_p)^Q\,{\bar G}\vp_Q(\{n\vp_j\},y_q^N/x_p^N).
\label{nhto}
\end{equation}
Next we use (\ref{epo2}) and (III.58) to derive
\begin{eqnarray}
\langle\Omega|\hat{\mathcal{T}}\vp_Q(y\vp_q,x\vp_q)
{\bf E}_{m,Q}^{+}|\Omega\rangle
=&z\vp_{m,Q}(\beta^Q_{m,0}/\Lambda_0^Q)N^{1-\frac 12L}x_p^{rN}
\nonumber\\
&\times(1-y_q^N/x_p^N)(y\vp_q/x\vp_p)^Q\,
{\bar{\mbox{\myeu h}}}^Q_{m}(y^N_q/x^N_p),
\label{ohteo}
\end{eqnarray}
where by (III.59) we have the polynomial identity
${\bar{\mbox{\myeu h}}}^Q_{m}(z)={\mbox{\myeu h}}^Q_{m}(z)$,
so that we can use (III.57), (\ref{roots}) and (\ref{beta0})
to evaluate the second ratio as
\begin{equation}
\frac{\langle\Omega|{\hat\mathcal{T}}\vp_Q(y\vp_q,x\vp_q)|\Omega\rangle}
{\langle\Omega|{\hat\mathcal{T}}\vp_Q(y\vp_q,x\vp_q)
{\bf E}_{m,Q}^{+}|\Omega\rangle}
=-\frac{x_p^N-y_q^N z_{m,Q}^{-1}}{x_p^N-y_q^N}
=\frac{{\bar X}\vp_{j,Q}+{\bar Y}\vp_{j,Q}}
{-z\vp_{m,Q}{\bar Z}\vp_{j,Q}}.
\label{ratio3}
\end{equation}
The $j$th factor in the product of (\ref{transfer2}) yields
\begin{equation}
\fl[{\bar X}\vp_{j,Q}-{\bf H}\vp_{j,Q}{\bar Y}\vp_{j,Q}+
({\bf E}^+_{j,Q}-z\vp_{j,Q}{\bf E}^-_{j,Q}){\bar Z}\vp_{j,Q}]
={\bf\cal R}\vp_{j,Q}
[{\bar A}\vp_{j,Q}-{\bf H}\vp_{j,Q}{\bar B}\vp_{j,Q}]
{\bf\cal S}_{j,Q}^{-1}.
\label{matrix}
\end{equation}
Therefore, as we chose the determinants of ${\bf \cal R}_{j,Q}$ and
${\bf \cal S}_{j,Q}$ to be one, we find
\begin{equation}
({\bar X}^2_{j,Q}-{\bar Y}^2_{j,Q}+z\vp_{j,Q}{\bar Z}^2_{j,Q})
=({\bar A}^2_{j,Q}-{\bar B}^2_{j,Q}).
\label{det}
\end{equation}
By inverting both sides of (\ref{matrix}), and using (\ref{det}),
we express (\ref{matrix}) in the diagonal representation
of ${\bf H}_{j,Q}$ as
\begin{equation}
\fl\left[\begin{array}{cc}
{\bar X}_{j,Q}+{\bar Y}_{j,Q}&-{\bar Z}_{j,Q}\\ 
z_{j,Q}{\bar Z}_{j,Q}&{\bar X}_{j,Q}-{\bar Y}_{j,Q}
\end{array}\right]
={\bf\cal S}\vp_{j,Q}
\left[\begin{array}{cc}
{\rm e}^{\theta_{j,Q}}-\lambda_q{\rm e}^{-\theta_{j,Q}}&0\\ 
0&{\rm e}^{-\theta_{j,Q}}-\lambda_q{\rm e}^{\theta_{j,Q}}
\end{array}\right]{\bf \cal R}_{j,Q}^{-1}.
\label{matrix2}
\end{equation}
Similarly the $j$th term in the product in (\ref{transfer1a})
can be written as
\begin{equation}
\fl\left[\begin{array}{cc}
X_{j,Q}-Y_{j,Q}& Z_{j,Q}\\ 
-z_{j,Q}Z_{j,Q} & X_{j,Q}+Y_{j,Q}\end{array}\right]
={\bf \cal S}\vp_{j,Q}
\left[\begin{array}{cc}
{\rm e}^{-\theta_{j,Q}}-\lambda_q^{-1}{\rm e}^{\theta_{j,Q}}&0\\ 
0&{\rm e}^{\theta_{j,Q}}-\lambda_q^{-1}{\rm e}^{-\theta_{j,Q}}
\end{array}\right]{\bf\cal R}_{j,Q}^{-1}.
\label{matrix1}
\end{equation}
It is easy to see that (\ref{ratio1}) gives the three elements of
the lower right triangle in the left hand side of (\ref{matrix1})
except for a constant factor ${\epsilon}_{j,Q}$, that is
\begin{eqnarray}
\fl X\vp_{j,Q}+Y\vp_{j,Q}
={\epsilon}\vp_{j,Q} k(y_p^N z\vp_{j,Q}-x_q^N)
={\epsilon}\vp_{j,Q}
[(1-k'\lambda\vp_p)z\vp_{j,Q}-(1-k'\lambda^{-1}_q)],
\nonumber\\
Z\vp_{j,Q} ={\epsilon}\vp_{j,Q} k(y_p^N -x_q^N)
={\epsilon}\vp_{j,Q} k'(\lambda^{-1}_q-\lambda\vp_p),
\end{eqnarray}
while (\ref{ratio3}) determines the upper left triangle in
(\ref{matrix2}) except for a constant ${\bar\epsilon}_{j,Q}$, i.e.
\begin{eqnarray}
\fl{\bar X}\vp_{j,Q}+{\bar Y}\vp_{j,Q}
={\bar\epsilon}\vp_{j,Q}k(x_p^N z\vp_{j,Q}-y_q^N)
={\bar\epsilon}\vp_{j,Q}
[(1-k'\lambda^{-1}_p)z\vp_{j,Q}-(1-k'\lambda\vp_q)],
\nonumber\\
{\bar Z}\vp_{j,Q}={\bar\epsilon}\vp_{j,Q} k(x_p^N -y_q^N)
={\bar\epsilon}\vp_{j,Q}k'(\lambda\vp_q-\lambda^{-1}_p).
\end{eqnarray}
The matrices in (\ref{matrix1}) are linear in $\lambda_q^{-1}$,
while those in (\ref{matrix2}) are linear in $\lambda\vp_q$.
Thus by equating the constant and linear terms, we find two
equations each for two matrices ${\bf M}$ and ${\bf N}$ defined
as in (II.90) for $Q=0$, namely
\begin{equation}
\fl{\bf\cal S}\vp_{j,Q}
\left[\begin{array}{cc}
{\rm e}^{-\theta_{j,Q}}&0\\ 
0&{\rm e}^{\theta_{j,Q}}
\end{array}\right]{\bf\cal R}_{j,Q}^{-1}={\bf M},\quad
{\bf\cal S}\vp_{j,Q}
\left[\begin{array}{cc}
{\rm e}^{\theta_{j,Q}}&0\\ 
0&{\rm e}^{-\theta_{j,Q}}
\end{array}\right]{\bf\cal R}_{j,Q}^{-1}=-{\bf N}.
\label{matrix3}
\end{equation}
Equations (\ref{matrix2}) and (\ref{matrix1}) are consistent, as
the diagonal elements also determine $X\vp_{j,Q}-Y\vp_{j,Q}$ and
${\bar X}\vp_{j,Q}-{\bar Y}\vp_{j,Q}$, whereas the off-diagonal
elements agree if one chooses
${\bar\epsilon}_{j,Q}=-{\epsilon}_{j,Q}\lambda_p$. Hence, all
matrix elements of $\bf M$ and $\bf N$ are explicitly found as
\begin{eqnarray}
\fl m_{11}={\epsilon}_{j,Q} k'\lambda_p,\,
m_{21}={\epsilon}_{j,Q}k'\lambda_p z_{j,Q},\,
m_{12}=-{\epsilon}_{j,Q}k'\lambda_p,\,
m_{22}={\epsilon}_{j,Q}(z_{j,Q}-1-k'z_{j,Q}\lambda_p),
\nonumber\\
\fl n_{11}=
{\epsilon}_{j,Q}(\lambda_pz_{j,Q}-\lambda_p-k'z_{j,Q}),\quad
n_{21}=-{\epsilon}_{j,Q}k'z_{j,Q},\quad
n_{12}=n_{22}={\epsilon}_{j,Q} k',
\label{mnij2}
\end{eqnarray}
which is a direct generalization of (\ref{mnij1}) to $Q\ne0$.
Evaluating the determinants of both sides of (\ref{matrix3}),
we again find ${\epsilon}^2_{j,Q} k'(z\vp_{j,Q}-1)\lambda\vp_p=1$.
Consequently, the matrices ${\bf\cal S}_{j,Q}$ and
${\bf\cal R}_{j,Q}$ can be evaluated in exactly the same way as
in \cite{APsu2} and subsection \ref{errors}, with the result,
[see also (II.92) and (II.93)],
\begin{eqnarray}
{\bf \cal S}\vp_{j,Q}=\half(s\vp_{11}+s\vp_{22}){\bf 1}
+\half(s\vp_{11}-s\vp_{22}){\bf H}\vp_{j,Q}
+s\vp_{12}{\bf E}^+_{j,Q}+s\vp_{21}{\bf E}^-_{j,Q},
\label{Sj}\\
{\bf \cal R}\vp_{j,Q}=\half(r\vp_{11}+r\vp_{22}){\bf 1}
+\half(r\vp_{11}-r\vp_{22}){\bf H}\vp_{j,Q}
+r\vp_{12}{\bf E}^+_{j,Q}+r\vp_{21}{\bf E}^-_{j,Q},
\end{eqnarray}
where, after fixing the free parameter $s\vp_{22}$ by the analog
of (\ref{Rij}),
\begin{eqnarray}
\fl s\vp_{22}=r\vp_{11}
=\biggl(\frac{m\vp_{22}\rme^{\theta_{j,Q}}
+n\vp_{22}\rme^{-\theta_{j,Q}}}
{2\sinh 2\theta\vp_{j,Q}}\biggr)^{\frac 12},\quad
&&
s\vp_{12}=z^{-1}_{j,Q}r\vp_{21}
=\frac{m\vp_{12}\rme^{\theta_{j,Q}}+n\vp_{12}\rme^{-\theta_{j,Q}}}
{m\vp_{22}\rme^{\theta_{j,Q}}+n\vp_{22}\rme^{-\theta_{j,Q}}}s\vp_{22},
\nonumber\\
\fl s\vp_{21}=z\vp_{j,Q}r\vp_{12}
=\frac{\rme^{-2\theta_{j,Q}}-k'}{2s\vp_{12}\sinh2\theta\vp_{j,Q}},\quad
&& s\vp_{11}=r\vp_{22}
=\frac{\rme^{2\theta_{j,Q}}-k'}{2s\vp_{22}\sinh2\theta\vp_{j,Q}}.
\label{Sij}
\end{eqnarray} 

%%%%%%%%%%%%%%%%%%%%%%%%%%%%%%%%%%%%%%%%%%%%%%%%%%%%%%%%%%%%%%%%%%%%%%%%
\subsection{Eigenvectors corresponding to
$y_q^{N-Q}\mathcal{G}^{}_b(\lambda^{}_q)$ }

We now consider eigenvectors of the transfer matrix related to
(\ref{eigv2}) and (\ref{gg3}). From (\ref{eigvec2}) and (\ref{C0B1m}),
we find that the generators of the corresponding ${\mathfrak{sl}}_2$
algebra can be written, similar to (\ref{ome}) and (\ref{epo}), as
\begin{eqnarray}
\langle\bar\Omega|{\bfE}_{m,Q}^{+}
=\omega^{Q(Q+1)}(\beta^Q_{m,0}/\Lambda^Q_{0})
\sum_{\ell=1}^{m_Q}z_{m,Q}^{\ell}
\langle\bar\Omega|{\bfB}_1^{(\ell N+Q)}{\bfC}_0^{(\ell N-N+Q)},
\label{bome}\\
{\bfE}_{m,Q}^-|\bar\Omega\rangle
=-\omega^{Q(Q+1)}(\beta^Q_{m,0}/\Lambda^Q_{0})
\sum_{\ell=1}^{m_Q}z_{m,Q}^{\ell-1}
{\bfB}_1^{(\ell N-N+Q)}{\bfC}_0^{(\ell N+Q)}|\bar\Omega\rangle,
\label{epbo}
\end{eqnarray}
which are generalizations of the second equations in (II.53)
and (II.54). Similar to the derivation of (\ref{ome2})
and (\ref{epo2}), we generalize (II.67) and (II.69) to
\begin{eqnarray}
\fl\langle\bar\Omega|{\bfE}_{m,Q}^{+}
=-(\beta^Q_{m,0}/\Lambda^Q_{0})z\vp_{m,Q}
\sum_{{\{0\le n_j\le N-1\}}\atop{n_1+\cdots+n_L=N}}
\langle\{N-1-n_j\}|\,\bar{G}^{\vp}_Q(\{n_j\},z^{\vp}_{m,Q}),
\label{ombe2}\\
\fl{\bfE}_{k,Q}^{-}|\bar\Omega\rangle
=(\beta^Q_{k,0}/\Lambda^Q_{0})
\sum_{{\{0\le n_j\le N-1\}}\atop{n_1+\cdots+n_L=N}}
{G}^{\vp}_Q(\{n_j\},z^{\vp}_{k,Q})|\{N-1-n_j\}\rangle.
\label{epbo2}
\end{eqnarray} 
Again, we use the theorem in \cite{APsu3} to find that 
\begin{equation}
\langle\bar \Omega|{\bfE}_{m,Q}^{+}
{\bfE}_{m,Q}^{-}|\bar\Omega\rangle=
\langle\bar \Omega|{\bfH}\vp_{m,Q}|\bar\Omega\rangle=1.
\label{bohbo}
\end{equation}
Comparing these results with those for ${\bf E}_{m,Q}^{\pm}$ in 
(\ref{ome2}) and (\ref{epo2}), we can see that what was done in
subsection \ref{loop} can be repeated here to obtain the generators
of a different quantum loop subalgebra, with generators
${\bfE}_{m,Q}^{\pm}$ as in (\ref{epem}) and ${\bfH}\vp_{m,Q}$ as
in (\ref{eH}).

Using (\ref{gg1}) and (\ref{eigent}), we may write (on the current
vector subspace)
\begin{eqnarray}
\fl{\cal T}\vp_Q(x\vp_q,y\vp_q)
&&=y_q^{N-Q}{\hat D}\vp_Q\prod_{j=1}^{m_Q}
\left[X^{\ast}_{j,Q}-{\bfH}\vp_{j,N-Q} Y^{\ast}_{j,Q}
+({\bfE}^+_{j,N-Q} -z^{\ast}_{j,Q}{\bfE}^-_{j,N-Q})Z^{\ast}_{j,Q}\right]
\nonumber\\
\fl&&=y_q^{N-Q}{\hat D}\vp_Q\prod_{j=1}^{m_Q}{\bf \cal S}^{\ast}_{j,Q}
(A^{\ast}_{j,Q}-{\bf H}\vp_{j,N-Q}B^{\ast}_{j,Q})
{\bf \cal R}^{\ast\,-1}_{j,Q},
\label{transfer3}
\end{eqnarray}
so that the $2^{m_Q}$ corresponding eigenvectors of the transfer matrix
are given by
\begin{equation}
\fl|\bar{\bf \cal X}^Q_s\rangle=\prod_{j=1}^{m_Q}{\bf \cal R}^{\ast}_{j,Q}
\prod_{m\in W_n}{\bfE}_{m,N-Q}^{-}|\bar\Omega\rangle,\quad
|\bar{\bf \cal Y}^Q_s\rangle=\prod_{j=1}^{m_Q}{\bf \cal S}^{\ast}_{j,Q}
\prod_{m\in W_n}{\bfE}_{m,N-Q}^{-}|\bar\Omega\rangle.
\label{evector2}
\end{equation}
Here $W_n=\{j_1,\ldots,j_n\}$, for $0\le n\le m_Q$, is the subset
of $\{1,2,\ldots,m_Q\}$, defined by $j\in W_n$ if $s_j=1$ and
$j\not\in W_n$ if $s_j=0$ otherwise. Using this notation we have
\begin{equation}
{\cal T}\vp_Q(x\vp_q,y\vp_q)|\bar{\bf\cal X}^Q_s\rangle=
y_q^{N-Q}{\hat D}\vp_Q
\prod_{j=1}^{m_Q}\left[A^{\ast}_{j,Q}-(-1)^{s_j}B^{\ast}_{j,Q}\right]
|\bar{\bf \cal Y}^Q_s\rangle.
\label{eig2}
\end{equation}
We may follow the procedure in subsection \ref{eigvec} to get
\begin{eqnarray}
\fl{\bf\cal S}^{\ast}_{j,Q}\;
=\half(s'_{11}+s'_{22}){\bf 1}
+\half(s'_{11}-s'_{22}){\bfH}\vp_{j,N-Q}
+s'_{12}{\bfE}^+_{j,N-Q}+s'_{21}{\bfE}^-_{j,N-Q},
\label{Spj}\\
\fl{\bf\cal R}^{\ast}_{j,Q}
=\half(r'_{11}+r'_{22}){\bf 1}
+\half(r'_{11}-r'_{22}){\bfH}\vp_{j,N-Q}
+r'_{12}{\bfE}^+_{j,N-Q}+r'_{21}{\bfE}^-_{j,N-Q},
\end{eqnarray} 
where the $s'_{ik}$ and $r'_{ik}$ are again given by (\ref{Sij}),
but with the replacements $z\vp_{j,Q}\to z^{\ast}_{j,Q}=z^{-1}_{j,Q}$
and $\theta_{j,Q}\to\theta^{\ast}_{j,Q}=\theta\vp_{j,N-Q}$.

%%%%%%%%%%%%%%%%%%%%%%%%%%%%%%%%%%%%%%%%%%%%%%%%%%%%%%%%%%%%%%%%%%%%%%%%

\section{Summary and Outlook}

The superintegrable transfer matrices have an Ising-like spectrum
\cite{BaxIf1} as shown in (\ref{gg1}) and (\ref{eigent}). In
(\ref{transfer1a}), the transfer matrices are expressed in terms of
the generators ${\bfE}^\pm_{j,Q}$, and ${\bfH}\vp_{j,Q}$ of
${\mathfrak{sl}}_2$ algebra.  These operate on the $N^{L-1}$
dimensional space of the edge variables $\{n_i\}$ satisfying the
cyclic condition $n_1+\cdots+n_L=0 ({\rm mod} N)$, but have the
${\mathfrak{sl}}_2$ commutation relations
\begin{equation}
[{\bf E}_{\ell,Q}^+,{\bf E}_{n,Q}^-]=
\delta\vp_{\ell,n}{\bf H}\vp_{\ell,Q},\quad
[{\bf H}\vp_{\ell,Q},{\bf E}_{n,Q}^\pm]=
\pm2\delta\vp_{\ell,n}{\bf E}_{\ell,Q}^\pm.
\label{commu}
\end{equation}
These operators are given in (\ref{epem}) and (\ref{H}) in terms
of the loop algebra defined by (\ref{xmpo}) and (\ref{xmphg})
satisfying (\ref{ome2}) and (\ref{epo2}). With the help of them
$2^{m_Q}$ eigenvectors of the transfer matrix (\ref{transfer1a})
are given in (\ref{evector}), where the elements of the $2\times2$
matrices $\bf \cal R$ and $\bf \cal S$ are explicitly given in
(\ref{Sij}).

In order to determine the complete set of eigenvectors we may have
to resort to a construction using both Bethe Ansatz methods
\cite{Tara,Roan} and methods from this paper. We do not need this
complication for the calculation of the order parameter and the pair
correlation of the superintegrable chiral Potts quantum chain in the
commensurate phase. These calculations can be done using the
eigenvectors presented here and we shall come back to this in later
works \cite{APsu5}.

%%%%%%%%%%%%%%%%%%%%%%%%%%%%%%%%%%%%%%%%%%%%%%%%%%%%%%%%%%%%%%%%%%%%%%%%

\section*{Acknowledgments}

We thank our colleagues and the staff at the Centre for Mathematics and
its Applications (CMA) and at the Department of Theoretical Physics
(RSPE) of Australian National University for their generous support and
hospitality.

%%%%%%%%%%%%%%%%%%%%%%%%%%%%%%%%%%%%%%%%%%%%%%%%%%%%%%%%%%%%%%%%%%%%%%%%

\appendix
\section{Identities (\ref{nxm}) and (\ref{mnxpxm})}

We start by generalizing (\ref{cbok}), using (\ref{C0B1m}), (II.55),
(III.7) and $\sum_j\bar N'_jn\vp_j=\sum_j n'_jN\vp_j$, as
\begin{equation}
\fl{\bfC}_0^{(mN+Q)}{\bfB}_1^{(nN+Q)}|\Omega\rangle
=\omega^{-Q}
\sum_{{\{0\le n_j\le N-1\}}\atop{n_1+\cdots+n_L=(n-m)N}}
\omega^{-\sum_{j}j n_j}K_{Nm+Q}(\{n_j\})|\{n_j\}\rangle,
\label{CBaux}
\end{equation}
valid for $n\ge m\ge0$.
To prove (\ref{nxm}) by induction, it is easily seen from (\ref{xm})
that it holds for $n=1$. We now assume this is also true for $n=m$,
so that
\begin{equation}
\fl({\bf x}^{-}_{1,Q})^m|\Omega\rangle
=\frac{m!\,\omega^Q}{\Lambda^Q_{0}}\,{\bfC}_0^{(Q)}
{\bfB}_1^{(m N+Q)} |\Omega\rangle
=\frac{m!}{\Lambda^Q_{0}}
\sum_{{\{0\le n_j\le N-1\}}\atop{n_1+\cdots+n_L=mN}}
\omega^{-\sum_{j}j n_j}K_Q(\{n_j\})|\{n_j\}\rangle,
\label{xmmo}
\end{equation}
after applying (\ref{CBaux}).
Again, we use (\ref{C0B1m}) and (II.55) to rewrite the action
of (\ref{xmpo}) on $|\{n_j\}\rangle$ with $\sum_j n_j=mN$ as
\begin{eqnarray}
\Lambda^Q_{0}{\bf x}^{-}_{1,Q}|\{n_j\}\rangle=
\sum_{{\{0\le\mu\vp_j,n'_j\le N-1\}}\atop{\sum n'_j=mN+N
\atop\sum\mu_j=Q\hfill}}
&&\omega^{\sum_{j}j(n\vp_j-n'_j)}\prod_{j=1}^L
\sfactor{n'_j+\mu\vp_j}{\mu_j}\sfactor{n'_j+\mu\vp_j}{n_j}
\nonumber\\
&&\times\,
\omega^{n\vp_j(N'_j+a\vp_j-N\vp_j)+\mu\vp_jN'_j}|\{n'_j\}\rangle,
\label{xmsm}
\end{eqnarray}
and
\begin{eqnarray}
\Lambda^Q_{0}{\bf x}^{+}_{0,Q}|\{n_j\}\rangle=
\sum_{{\{0\le\mu\vp_j,n'_j\le N-1\}}\atop{\sum n'_j=mN-N
\atop\sum \mu_j=N+Q\hfill}}
&&\omega^{\sum_{j}j(n\vp_j-n'_j)}\prod_{j=1}^L
\sfactor{n'_j+\mu\vp_j}{\mu\vp_j}\sfactor{n'_j+\mu\vp_j}{n\vp_j}
\nonumber\\
&&\times\,
\omega^{n\vp_j(N'_j+a\vp_j-N\vp_j)+\mu\vp_jN'_j}|\{n'_j\}\rangle.
\label{xpsm}
\end{eqnarray}
Here, as in our previous papers, we have defined
\begin{equation}
N\vp_j\equiv\sum_{\ell<j}n\vp_{\ell},\quad
N'_j\equiv\sum_{\ell<j}n'_{\ell},
\quad a\vp_j\equiv\sum_{\ell<j}\mu\vp_{\ell}.
\end{equation}
After multiplying (\ref{xmmo}) by ${\bf x}^-_{1,Q}$ and using
(\ref{xmsm}) together with (III.7) and (III.21), we find
\begin{eqnarray}
\fl({\bf x}^{-}_{1,Q})^{m+1}|\Omega\rangle
&&=\frac{m!}{(\Lambda^Q_{0})\vf^2}
\sum_{{\{0\le\lambda\vp_j,\mu\vp_j,n'_j\le N-1\}}
\atop{\sum n'_j=mN+N\atop\sum\mu_j=\sum\lambda_j=Q}}
\omega^{-\sum_{j}j n'_j}
I_{mN}(\{n'_j+\mu\vp_j\};\{\lambda\vp_j\})
\nonumber\\
&&\hspace{11em}\times\,\prod_{j=1}^L\sfactor{n'_j+\mu\vp_j}{\mu\vp_j}
\omega^{\mu\vp_jN'_j}|\{n'_j\}\rangle
\nonumber\\
&&=\frac{(m+1)!}{\Lambda^Q_{0}}
\sum_{{\{0\le n'_j\le N-1\}}\atop{\sum n'_j=mN+N}}
\omega^{-\sum_{j}j n'_j}K_Q(\{n'_j\})|\{n'_j\}\rangle
\nonumber\\
&&=\frac{(m+1)!\,\omega^Q}{\Lambda^Q_{0}}\,
{\bfC}_0^{(Q)}{\bfB}_1^{(m N+N+Q)} |\Omega\rangle,
\label{xmm1o}
\end{eqnarray}
where (III.22) of Lemma 1 in \cite{APsu3} is used and also
(III.7) and (\ref{CBaux}) to carry out the other two sums.
This then proves (\ref{nxm}). 
To prove (\ref{mnxpxm}), we first prove it for $m=1$. After
multiplying (\ref{xmmo}) (in which $m$ is replaced by $n$)
by ${\bf x}^+_{0,Q}$, and then using (\ref{xpsm}), we get
\begin{eqnarray}
\fl{\bf x}^{+}_{0,Q}({\bf x}^{-}_{1,Q})^{n}|\Omega\rangle
&&=\frac{n!}{(\Lambda^Q_{0})\vf^2}
\sum_{{\{0\le\lambda\vp_j,\mu\vp_j,n'_j\le N-1\}}
\atop{\sum n'_j=nN-N\atop\sum\lambda_j=Q,\;\sum \mu_j=N+Q}}
\omega^{-\sum_{j}j n'_j}
I_{nN}(\{n'_j+\mu\vp_j\};\{\lambda\vp_j\})
\nonumber\\
&&\hspace{11em}\times\,\prod_{j=1}^L\sfactor{n'_j+\mu\vp_j}{\mu\vp_j}
\omega^{\mu\vp_jN'_j}|\{n'_j\}\rangle.
\end{eqnarray}
Again we use (III.22) of Lemma 1 in \cite{APsu3}, then use
(III.7) and (\ref{CBaux}) to obtain
\begin{eqnarray}
{\bf x}^{+}_{0,Q}({\bf x}^{-}_{1,Q})^{n}|\Omega\rangle
&&=\frac {n!}{\Lambda^Q_{0}}
\sum_{{\{0\le n'_j\le N-1\}}\atop{\sum n'_j=nN-N}}
\omega^{-\sum_{j}j n'_j}K_{N+Q}(\{n'_j\})|\{n'_j\}\rangle
\nonumber\\
&&=\frac{n!\,\omega^Q}{\Lambda^Q_{0}}\,{\bfC}_0^{(N+Q)}
{\bfB}_1^{(n N+Q)} |\Omega\rangle.
\label{1nxpxm}
\end{eqnarray}
This shows that (\ref{mnxpxm}) holds for $m=1$. Next assume
that (\ref{mnxpxm}) holds for $m=\ell$, so that from (\ref{CBaux})
\begin{equation}
({\bf x}^{+}_{0,Q})^\ell({\bf x}^{-}_{1,Q})^n|\Omega\rangle
=\frac{\ell!n!}{\Lambda^Q_{0}}
\sum_{{\{0\le n_j\le N-1\}}\atop{\sum n_j=nN-\ell N}}
\omega^{-\sum_{j}j n_j}K_{\ell N+Q}(\{n_j\})|\{n_j\}\rangle.
\label{lnxpxm}
\end{equation}
To prove that it also holds for $m=\ell+1$, we multiply
(\ref{lnxpxm}) by ${\bf x}^+_{0,Q}$ and then use (\ref{xpsm}),
together with (III.7) and (III.21)\footnote{In (III.21) we
identify $\sum_j n_j\bar b_j=\sum_j\lambda_j N_j$, as
follows from definitions (III.7) and (III.20).}, to find
\begin{eqnarray}
&\frac{(\Lambda^Q_{0})_{\vp}^2}{\ell!n!}&
({\bf x}^{+}_{0,Q})^{\ell+1}({\bf x}^{-}_{1,Q})^n|\Omega\rangle
\nonumber\\
&&=\sum_{{\{0\le\lambda\vp_j,\mu\vp_j,n'_j\le N-1\}}
\atop{\sum n'_j=nN-\ell N-N\atop\sum\lambda_j=\ell N+Q,\;\sum\mu_j=N+Q}}
\omega^{-\sum_{j}j n'_j}I_{nN-\ell N}(\{n'_j+\mu\vp_j\};\{\lambda\vp_j\})
\nonumber\\
&&\hspace{9em}\times\,\prod_{j=1}^L\sfactor{n'_j+\mu\vp_j}{\mu\vp_j}
\omega^{\mu\vp_jN'_j}|\{n'_j\}\rangle\
\nonumber\\
&&=\sum_{{\{0\le\lambda\vp_j,\mu\vp_j,n'_j\le N-1\}}
\atop{\sum n'_j=nN-\ell N-N\atop\sum\lambda_j=\ell N+Q,\;\sum\mu_j=N+Q}}
\omega^{-\sum_{j}j n'_j}
\bI_{\ell N}(\{\lambda\vp_j\};\{n'_j+\mu\vp_j\})
\nonumber\\
&&\hspace{9em}\times\,\prod_{j=1}^L\sfactor{n'_j+\mu\vp_j}{\mu\vp_j}
\omega^{\mu\vp_jN'_j}|\{n'_j\}\rangle,
\label{la1nxpxm}
\end{eqnarray}
where (III.24) is used. Using (III.26) and the identities,
\begin{equation}
\fl\sfactor{n'_j+\mu\vp_j}{\mu\vp_j}
\sfactor{n'_j+n\vp_j+\mu\vp_j}{n\vp_j}
=\sfactor{n\vp_j+\mu\vp_j}{n\vp_j}
\sfactor{n'_j+n\vp_j+\mu\vp_j}{n\vp_j+\mu\vp_j},
\quad \sum_{j=1}^L n_j=\ell N,
\label{rsfactor}
\end{equation}
we may rewrite (\ref{la1nxpxm}), by making the change of variables
$\mu\vp_j=\mu'_j-n\vp_j$ and $\lambda\vp_j=\lambda'_j+n\vp_j$
(with $\sum\mu'_j=\ell N+N +Q$ and $\sum\lambda'_j=Q$) followed
by applying (III.21), as
\begin{eqnarray}
\fl&\frac{(\Lambda^Q_{0})_{\vp}^2}{\ell!n!}
&({\bf x}^{+}_{0,Q})^{\ell+1}({\bf x}^{-}_{1,Q})^n|\Omega\rangle
\nonumber\\
\fl
&&=\sum_{{\{0\le\lambda'_j,\mu'_j,n'_j\le N-1\}}
\atop{\sum n'_j=nN-\ell N-N\atop
\sum\lambda'_j=Q,\;\sum\mu'_j=\ell N+N+Q}}
\omega^{-\sum_{j}j n'_j}
I_{\ell N}(\{\mu'_j\};\{\lambda'_j\})
\nonumber\\
&&\hspace{9em}\times\,\prod_{j=1}^L\sfactor{n'_j+\mu'_j}{\mu'_j}
\omega^{\mu\vp_jN'_j}|\{n'_j\}\rangle\label{inter}\\
\nonumber\\
&&=(\ell+1)\Lambda^Q_{0}
\sum_{{\{0\le n'_j\le N-1\}}\atop{\sum n'_j=nN-\ell N-N}}
\omega^{-\sum_{j}j n'_j}K_{\ell N+N+Q}(\{n'_j\})|\{n'_j\}\rangle
\nonumber\\
&&=(\ell+1)\omega\vn^Q\Lambda^Q_{0}\,{\bfC}_0^{(\ell N+N+Q)}
{\bfB}_1^{(nN+Q)} |\Omega\rangle.
\label{l1nxpxm}
\end{eqnarray}
In (\ref{inter}), (III.22), (III.7) and (\ref{CBaux}) are to be used
again to arrive at (\ref{l1nxpxm}). This proves that (\ref{mnxpxm})
holds for $m=\ell+1$, and therefore holds for all $m$.

%%%%%%%%%%%%%%%%%%%%%%%%%%%%%%%%%%%%%%%%%%%%%%%%%%%%%%%%%%%%%%%%%%%%%%%%
\section{Serre Relation for Special Cases}

Let $\ell=1$ in (\ref{lnxpxm}), and multiply it by ${\bf x}^-_{1,Q}$;
next use (III.7) and (\ref{xmsm}), followed by (III.21) to obtain
\begin{eqnarray}
\fl(\Lambda^Q_{0})_{\vp}^2{\bf x}^-_{1,Q}({\bf x}^{+}_{0,Q})
({\bf x}^{-}_{1,Q})^n|\Omega\rangle
=n!\sum_{{\{0\le\lambda\vp_j,\mu\vp_j,n'_j\le N-1\}}
\atop{\sum n'_j=nN,\atop\sum\lambda_j=N+Q,\sum\mu_j=Q}}
\omega^{-\sum_{j}j n'_j}
I_{nN-N}(\{n'_j+\mu\vp_j\};\{\lambda\vp_j\})
\nonumber\\
\hspace{14em}\times\,\prod_{j=1}^L\sfactor{n'_j+\mu\vp_j}{\mu\vp_j}
\omega^{\mu\vp_jN'_j}|\{n'_j\}\rangle,
\label{xmxpxmn}
\end{eqnarray}
following analogous steps as in the derivation of (\ref{la1nxpxm}).
Using (III.25) with $n\to1$, $\ell\to n$ and
$\bI_0(\{\lambda_j\};\{\mu_j\})=1$, we find
\begin{equation}
I_{n N-N}(\{\mu\vp_j+n'_j\};\{\lambda\vp_j\})
=1+(n-1)\bI_N(\{\lambda\vp_j\};\{\mu\vp_j+n'_j\}).
\label{IbI}
\end{equation}
Similar to the derivation of (\ref{inter}) from (\ref{la1nxpxm}),
we use (III.26), (\ref{rsfactor}) and (III.21), changing variables
$\mu\vp_j=\mu'_j-n\vp_j$ and $\lambda\vp_j=\lambda'_j+n\vp_j$, to find
\begin{eqnarray}
\fl\sum_{{\{0\le\mu_j,\lambda_j\le N-1\}}
\atop{\sum\lambda_j=N+Q,\sum \mu_j=Q}}
\bI_{N}(\{\lambda\vp_j\};\{n'_j+\mu\vp_j\})
\prod_{j=1}^L\sfactor{n'_j+\mu\vp_j}{\mu\vp_j}\omega^{\mu\vp_jN'_j}
\nonumber\\
\fl\qquad=\sum_{{\{0\le\mu'_j,\lambda'_j\le N-1\}}
\atop{\sum\lambda'_j=Q,\sum \mu'_j=N+Q}}
I_{N}(\{\mu'_j\};\{\lambda'_j\})
\prod_{j=1}^L\sfactor{n'_j+\mu'_j}{\mu'_j}\omega^{\mu'_jN'_j}
=\Lambda_0^Q K_{N+Q}(\{n_j'\}),
\label{IbI2}
\end{eqnarray}
where (III.22) and (III.7) have been used for the last equality.
Substituting (\ref{IbI}) into (\ref{xmxpxmn}), and using
(\ref{IbI2}), (III.22), (III.7) and also (\ref{roots}),
$\sum_{\{\lambda_j\},\sum\lambda_j=mN+Q}1=\Lambda_m^Q$,
we find
\begin{eqnarray}
\fl{\bf x}^-_{1,Q}{\bf x}^{+}_{0,Q}
({\bf x}^{-}_{1,Q})^n|\Omega\rangle
&&=\frac{n!}{(\Lambda^Q_{0})\vf^2}
\sum_{{\{0\le n'_j\le N-1\}}\atop{\sum n'_j=nN}}
\omega^{-\sum_{j}j n'_j}\bigg[\Lambda^Q_{1}K_Q(\{n'_j\})
\nonumber\\
&&\hspace{7em}
+(n-1)\Lambda_0^Q K_{N+Q}(\{n_j'\})\bigg]|\{n'_j\}\rangle
\nonumber\\
&&=\Bigg[\frac{\Lambda^Q_{1}}{\Lambda^Q_{0}}
({\bf x}^{-}_{1,Q})^n+\frac{n-1}{n+1}\,
{\bf x}^{+}_{0,Q}({\bf x}^{-}_{1,Q})^{n+1}\Bigg]|\Omega\rangle,
\label{serre1}
\end{eqnarray}
where (\ref{lnxpxm}) is used to get the second equality.
Multiplying (\ref{serre1}) by ${\bf x}^-_{1,Q}$ on both sides, we find
\begin{eqnarray}
\fl({\bf x}^-_{1,Q})^2\,{\bf x}^{+}_{0,Q}
({\bf x}^{-}_{1,Q})^n|\Omega\rangle
=\Bigg[
\frac{\Lambda^Q_{1}}{\Lambda^Q_{0}}
({\bf x}^{-}_{1,Q})^{n+1}+\frac{n-1}{n+1}\,
{\bf x}^-_{1,Q}{\bf x}^{+}_{0,Q}
({\bf x}^{-}_{1,Q})^{n+1}\Bigg]|\Omega\rangle
\nonumber\\
=\Bigg[\frac{2n}{n+1}
\frac{\Lambda^Q_{1}}{\Lambda^Q_{0}}
({\bf x}^{-}_{1,Q})^{n+1}+\frac{n(n-1)}{(n+1)(n+2)}\,
{\bf x}^{+}_{0,Q}({\bf x}^{-}_{1,Q})^{n+2}\Bigg]|\Omega\rangle,
\label{serre2}
\end{eqnarray}
where (\ref{serre1}) is used again for the second term, and the
coefficients are collected. Similarly, we can show
\begin{equation}
\fl({\bf x}^-_{1,Q})^3\,{\bf x}^{+}_{0,Q}
({\bf x}^{-}_{1,Q})^n|\Omega\rangle
=\Bigg[\frac{3n}{n+2}
\frac{\Lambda^Q_{1}}{\Lambda^Q_{0}}
({\bf x}^{-}_{1,Q})^{n+2}+\frac{n(n-1)}{(n+2)(n+3)}\,
{\bf x}^{+}_{0,Q}({\bf x}^{-}_{1,Q})^{n+3}\Bigg]|\Omega\rangle.
\label{serre3}
\end{equation}
By substituting (\ref{serre1}), (\ref{serre2}) and (\ref{serre3})
into the second member of the following equation and collecting terms
we find
\begin{eqnarray}
\fl[[[{\bf x}^{+}_{0,Q},{\bf x}^{-}_{1,Q}],{\bf x}^{-}_{1,Q}],
{\bf x}^{-}_{1,Q}]({\bf x}^-_{1,Q})^{n}|\Omega\rangle
=\{{\bf x}^{+}_{0,Q}({\bf x}^{-}_{1,Q})^{3+n}
-3{\bf x}^-_{1,Q}{\bf x}^{+}_{0,Q}
({\bf x}^{-}_{1,Q})^{2+n}
\nonumber\\
+3({\bf x}^{-}_{1,Q})^2\,{\bf x}^{+}_{0,Q}({\bf x}^-_{1,Q})^{1+n}
-({\bf x}^{-}_{1,Q})^3\,{\bf x}^{+}_{0,Q}
({\bf x}^{-}_{1,Q})^n\}|\Omega\rangle=0.
\label{serre4}
\end{eqnarray}

%%%%%%%%%%%%%%%%%%%%%%%%%%%%%%%%%%%%%%%%%%%%%%%%%%%%%%%%%%%%%%%%%%%%%%%%


\section*{References}
\begin{thebibliography}{000}

%
\bibitem{AMPTY}{%
Au-Yang H, McCoy B M, Perk J H H, Tang S and Yan M-L 1987
\textrm{Commuting transfer matrices in the chiral Potts models:
Solutions of the star-triangle equations with genus $> 1$}
\textit{Phys. Lett.} A {\bf 123} 219--23}
%
\bibitem{BPAuY88}{%
Baxter R J, Perk J H H and Au-Yang H 1988
\textrm{New solutions of the star-triangle relations for the chiral
Potts model}
\textit{Phys. Lett.} A {\bf 128} 138--42}
%
\bibitem{vonGehlen1985}{%
von Gehlen G and Rittenberg V 1985
\textrm{$Z_n$-symmetric quantum chains with an infinite set of
conserved charges and $Z_n$ zero modes}
\textit{Nucl. Phys. B} {\bf 257} 351--70}
\bibitem{Baxsu}{%
Baxter R J 1989
\textrm{Superintegrable chiral Potts model: Thermodynamic
properties, an ``inverse" model, and a simple associated Hamiltonian}
\textit{J. Stat. Phys.} {\bf 57} 1--39}
%
\bibitem{BaxIf1}{%
Baxter R J 1993
\textrm{Chiral Potts model with skewed boundary conditions}
\textit{J. Stat. Phys.} {\bf 73} 461--95}
%
\bibitem{BaxIf2}{%
Baxter R J 1994
\textrm{Interfacial tension of the chiral Potts model}
\textit{J. Phys. A: Math. Gen.} {\bf 27} 1837--49}
%
\bibitem{Tara}{%
Tarasov V O 1990
\textrm{Transfer matrix of the superintegrable chiral Potts model.
Bethe ansatz spectrum,}
\textit{Phys. Lett.} A {\bf 147} 487--90}
%
\bibitem{NiDe1}{% 
Nishino A and Deguchi T 2006
\textrm{The $L({\mathfrak sl}_2)$ symmetry of the Bazhanov--Stroganov
model associated with the superintegrable chiral Potts model}
\textit{Phys. Lett.} A {\bf 356} 366--70
\textit{Preprint} arXiv:cond-mat/0605551}
%
\bibitem{APsu1}{%
Au-Yang H and Perk J H H 2008
\textrm{Eigenvectors in the superintegrable model I:
${\mathfrak{sl}}_2$ generators}
\textit{J. Phys. A: Math. Theor.} {\bf 41} 275201 (10pp)
\textit{Preprint} arXiv:0710.5257}
%
\bibitem{APsu2}{%
Au-Yang H and Perk J H H 2009
\textrm{Eigenvectors in the superintegrable model II:
Ground state sector}
\textit{J.~Phys. A: Math. Theor.} {\bf 42} 375208 (16pp)
\textit{Preprint} arXiv:0803.3029}
%
\bibitem{DFM}{% 
Deguchi T, Fabricius K and McCoy B M 2001
\textrm{The $sl_2$ loop algebra symmetry of the six-vertex model
at roots of unity}
\textit{J. Stat. Phys.} {\bf 102} 701--36
\textit{Preprint} arXiv:cond-mat/9912141}
%
\bibitem{NiDe2}{% 
Nishino A and Deguchi T 2008
\textrm{An algebraic derivation of the eigenspaces associated with
an Ising-like spectrum of the superintegrable chiral Potts model}
\textit{J. Stat. Phys.} {\bf 133} 587--615
\textit{Preprint} arXiv:0806.1268}
%
\bibitem{Roan}{%
Roan S S 2010
\textrm{Eigenvectors of an arbitrary Onsager sector in
superintegrable $\tau^{(2)}$-model and chiral Potts model}
\textit{Preprint} arXiv:1003.3621}
%
\bibitem{Baxter-tau}{%
Baxter R J 2004
\textrm{Transfer matrix functional relations for the
generalized $\tau_2(t_q)$ model}
\textit{J. Stat. Phys.} {\bf 117} 1--25
\textit{Preprint} arXiv:cond-mat/0409493}
%
\bibitem{BBP}{%
Baxter R J, Bazhanov V V and Perk J H H 1990
\textrm{Functional relations for transfer matrices of the
chiral Potts model}
\textit{Int. J. Mod. Phys.} B {\bf 4} 803--70}
%
\bibitem{JimboNK}{%
Jimbo M 1992
\textrm{Topics from representations of $U_q(\mathfrak g)$ ---
an introductory guide to physicists}
\textit{Quantum Groups and Quantum Integrable Systems
(Nankai Lectures on Mathematical Physics)}
ed M-L Ge (Singapore: World Scientific) pp~1--61}
%
\bibitem{Degu3}{%
Deguchi T 2007
\textrm{Extension of a Borel subalgebra into the ${\mathit sl}_2$
loop algebra symmetry for the twisted XXZ spin chain at roots of
unity and the Onsager algebra}
\textit{Proc. Workshop on Recent Advances
in Quantum Integrable Systems (RAQIS'07)}
ed L Frappat and E Ragoucy
(Annecy-le-Vieux, France: LAPTH) pp~15--34
\textit{Preprint} arXiv:0712.0066}
%
\bibitem{APsu3}{%
Au-Yang H and Perk J H H 2010
\textrm{Identities in the superintegrable chiral Potts model}
\textit{J. Phys. A: Math. Theor.} {\bf 43} 025203 (10pp)
\textit{Preprint} arXiv:0906.3153}
%
\bibitem{ItoTer}{%
Ito T and P. Terwilliger P 2004
\textrm{The shape of a tridiagonal pair}
\textit{J. Pure Appl. Algebra} {\bf 188} 145--60
\textit{Preprint} arXiv:math/0304244}
%
\bibitem{Degu2}{%
Deguchi T 2007
\textrm{Irreducibility criterion for a finite-dimensional highest
weight representation of the ${\mathit sl}_2$ loop algebra
and the dimensions of reducible representations}
\textit{J. Stat. Mech.} P05007 1--30
\textit{Preprint} arXiv:math-ph/0610002}
%
\bibitem{Davies}{%
Davies B 1990
\textrm{Onsager's algebra and superintegrability}
\textit{J. Phys. A: Math. Gen.} {\bf 23} 2245--61}
%
\bibitem{APsu5}{%
Au-Yang H and Perk J H H 2010
\textrm{Spontaneous magnetization of the integrable
chiral Potts model}
\textit{Preprint} arXiv:1003.4805}
%
\end{thebibliography}
\end{document}